\documentclass[floats,floatfix,twocolumn,superscriptaddress,nofootinbib,showpacs]{revtex4-1}

\pdfoutput=1
\usepackage{amsfonts,amssymb,amsmath}

\usepackage[usenames,dvipsnames]{xcolor}
\usepackage[unicode,colorlinks=true,linkcolor=blue,urlcolor=blue,citecolor=gray]{hyperref}
\usepackage[all]{hypcap}
\usepackage{graphicx}
\usepackage{xspace}
\usepackage{tikz}
\usetikzlibrary{shapes,arrows}
\usetikzlibrary{matrix}
\usetikzlibrary{shapes.multipart}
\usetikzlibrary{er,positioning}
\usepackage[normalem]{ulem} 
\usepackage{url}
\usepackage{color,soul}
\usepackage[caption=false]{subfig}
\usepackage[T1]{fontenc}
\usepackage[utf8]{inputenc}
\usepackage{microtype}
\usepackage{float}

\newcommand{\Flatiron}{\affiliation{Center for Computational Astrophysics, Flatiron Institute, 162 5th Ave, New York, NY 10010}}

\begin{document}

\title{Revisiting non-linearity in binary black hole mergers}

\author{Maria Okounkova}
\email{mokounkova@flatironinstitute.org}
\Flatiron

\date{\today}

\begin{abstract}
Recently, it has been shown that with the inclusion of overtones, the post-merger gravitational waveform at infinity of a binary black hole system is well-modelled using pure linear theory. However, given that a binary black hole merger is expected to be highly non-linear, where do these non-linearities, which do not make it out to infinity, go? We visualize quantities measuring non-linearity in the strong-field region of a numerical relativity binary black hole merger in order to begin to answer this question.
\end{abstract}

\maketitle

\section{Introduction}
\label{sec:introduction}

Recently, it has been found that the post-merger gravitational radiation from a numerical relativity binary black hole merger can be fully described by linear theory~\cite{Giesler:2019uxc, Isi:2019aib}. This phenomenon has additionally been probed experimentally, using gravitational wave data from GW150914~\cite{Isi:2019aib}. This is a surprising result, given the highly non-linear nature of binary black hole mergers. Given that, at future null infinity, the gravitational radiation is purely linear, what happens to the non-linearities from merger? 

In a numerical relativity simulation, we have access not only to the gravitational waveform of a binary black hole merger, but the entire strong-field region through inspiral, merger, and ringdown. In this study, we thus attempt to measure non-linearities in the strong-field region, and track where these non-linearities go during the post-merger phase.

Specifically, we use a set of gauge-invariant quantities that can be evaluated pointwise on the computational domain that were first theoretically proposed in~\cite{lobo16} and whose properties we numerically explored in~\cite{Bhagwat:2017tkm}. These quantities measure the closeness of a spacetime to Kerr, and hence we refer to them as \textit{Kerrness} measures. In~\cite{Bhagwat:2017tkm}, we showed the ability of these measures to quantify non-linearities in black hole spacetimes (cf. Fig. 3 of~\cite{Bhagwat:2017tkm}), a key result that we will make use of in this study. 

We will use these Kerrness measures to track the behavior of non-linearities in the strong-field region of a numerical binary black hole merger, with the main `product' of this analysis being visualizations of these quantities on the computational domain as a function of time to see where the non-linearities `go'. 

We first introduce these measures of non-linearity in Sec.~\ref{sec:methods}, with more details available in~\cite{Bhagwat:2017tkm}. In Sec.~\ref{sec:results}, we present our results for investigating non-linearities on an equal mass, non-spinning binary black hole merger. In Sec.~\ref{sec:overtones}, we corroborate the results of~\cite{Giesler:2019uxc} by showing that the post-peak gravitational radiation is fully describable by linear perturbation theory. In Sec.~\ref{sec:nonlinear}, we investigate the behavior of the Kerrness measures in the strong field region. We summarize in Sec.~\ref{sec:discussion}. 

We set $G = c = 1$ throughout. Quantities are given in terms of units of $M$, the sum of the Christodoulou masses of the background black holes at a given relaxation time~\cite{Boyle:2009vi}. Throughout this paper, we will make use of some numerical relativity terms, mostly the concept of a spatial spacetime slices that arise from a 3+1 decomposition. For more details on the spatial slicing of spacetime, please see~\cite{baumgarteShapiroBook}.

\section{Methods}
\subsection{Measuring non-linearity}
\label{sec:methods}

In order to quantify non-linearities on the computational domain during a binary black hole merger, we will use a set of gauge-invariant, point-wise evaluated \textit{Kerrness measures}. These measures were first outlined theoretically in~\cite{lobo16}, and evaluate analytically to zero if and only if a spacetime is Kerr or a linear perturbation of Kerr. In~\cite{Bhagwat:2017tkm} we performed numerical computations with these measures that were not possible to be performed analytically. 

While the technical details and origins of these Kerrness measures are given in detail in~\cite{Bhagwat:2017tkm}, the \textit{key result} that we will make use of here is the ability of these measures to quantify non-linearities in a black hole spacetime. In~\cite{Bhagwat:2017tkm} we performed the following experiment. We began with a Kerr metric, $g_\mathrm{Kerr}$ (with indices dropped for ease of notation), and added a linear quasi-normal mode perturbation $h_\mathrm{QNM}$ (cf.~\cite{Yang:2014tla, Lousto:2002em, Teukolsky:2014vca}) with strength $\varepsilon$, giving a linearly perturbed metric
\begin{align}
\label{eq:linear_metric}
    g_\mathrm{linear} = g_\mathrm{Kerr} + \varepsilon h_\mathrm{QNM}\,.
\end{align}
The linearly-perturbed metric in Eq.~\eqref{eq:linear_metric} does \textit{not} fully satisfy the non-linear constraint equations (which arise from the Einstein field equations, cf.~\cite{baumgarteShapiroBook}), rather having errors at $\mathcal{O}(\varepsilon^2)$. 

We then \textit{re-solved} the non-linear constraint equations given the metric in Eq.~\eqref{eq:linear_metric} (cf.~\cite{Cook2004, Pfeiffer:2005zm, baumgarteShapiroBook}) to obtain a \textit{non-linearly perturbed} black hole spacetime, with non-linearities entering in at $\mathcal{O}(\varepsilon^2)$. 

We then evaluated the Kerrness measures on this non-linearly perturbed spacetime, and the results are given in Fig. 3 of~\cite{Bhagwat:2017tkm}. We found that as a function of the perturbation strength $\varepsilon$, the Kerrness measures increased \textit{quadratically} from zero for low values of $\varepsilon$. This in turn means that the Kerrness measures were picking up on non-linearities in the spacetime. At higher values of $\varepsilon$, the Kerrness measures increased with higher powers (cubic, quartic), showing that they were picking up on higher-order non-linearities. This is the main result from~\cite{Bhagwat:2017tkm} that we will make use of in this study - the ability of the Kerrness measures to quantify non-linearities in a black hole spacetime.

How does this result apply to the \textit{binary} black hole spacetimes we aim to consider in this study? Initially, at infinite separation, a binary black hole spacetime can be treated as a linear superposition of two Kerr spacetimes~\cite{Lovelace:2008hd}. Thus, we expect no non-linearities to be present, and expect the Kerrness measures to be zero. The final state of a binary black hole merger is a stationary Kerr black hole~\cite{Owen:2009sb, Campanelli:2008dv, Bhagwat:2017tkm}, on which the Kerrness measures are also zero. The merger process, however, introduces non-linearities in the spacetime surrounding the two black holes, as they can be thought of non-linearly `perturbing' one another. Thus, we start from a spacetime where the Kerrness measures are zero, end with a spacetime where the Kerrness measures are zero, and between the two will use the Kerrness measures to observe the formation of non-linearities and see where they go.

\subsubsection{Numerical details}
\label{sec:numerical_details}

Before we use these Kerrness measures to track non-linearities on the binary black hole merger spacetime, however, there is a numerical subtlety we must address. Theoretically, each of the Kerrness measures evaluates to precisely zero on a Kerr spacetime. However, in numerical practice, due to finite numerical resolution, these quantities are non-zero on a Kerr spacetime, but \textit{converge to zero} with increasing numerical resolution. Moreover, given that we use a pseudo-spectral code as detailed in Sec.~\ref{sec:simulation}, the Kerrness measures converge \textit{exponentially} to zero. We show a test of this convergence in Fig. 5 of~\cite{Bhagwat:2017tkm}. 

In a binary black hole simulation with finite numerical resolution, there is thus a numerical noise floor for the Kerrness measures, that is unphysical and due purely to numerical resolution. How can we correct for such noise when analyzing the behavior of the Kerrness measures? We know from multiple different diagnostics that the final remnant of a binary black hole simulation is a Kerr spacetime~\cite{Owen:2009sb, Campanelli:2008dv, Bhagwat:2017tkm}. We can thus use the non-zero values of the Kerrness quantities late after merger to infer the value of the numerical noise floor. Additionally, we use a combination of superposed Kerr-Schild black holes for the initial data, following~\cite{Lovelace:2008hd}. While we do re-solve the non-linear constraints for the initial, at large enough initial separation, the spacetime should be close to a linear superposition. We can thus use this to inform the numerical floor for the Kerrness quantities as well. 

Thus, when we investigate the Kerrness measures on the binary black hole merger, we will use this inferred numerical resolution floor as our `zero'. In other words, when we show a 2-dimensional colormap of a given Kerrness measure, instead of setting the lower bound of the colormap to $0.0$, we will instead set it to the numerical noise floor. The upper bound of the colormap will be set by the maximum value of the Kerrness quantity over time and the computational domain. These bounds will differ for each of the Kerrness quantities, are we are most interested in the \textit{relative} scale, in terms of how many orders of magnitude are spanned in non-linearity above the lower bound. As we shall see in Sec.~\ref{sec:nonlinear} will result in a range of $\sim 10^{5}$.

\subsection{Considering horizons}
\label{sec:horizons}

\begin{figure}
  \includegraphics[width=\columnwidth]{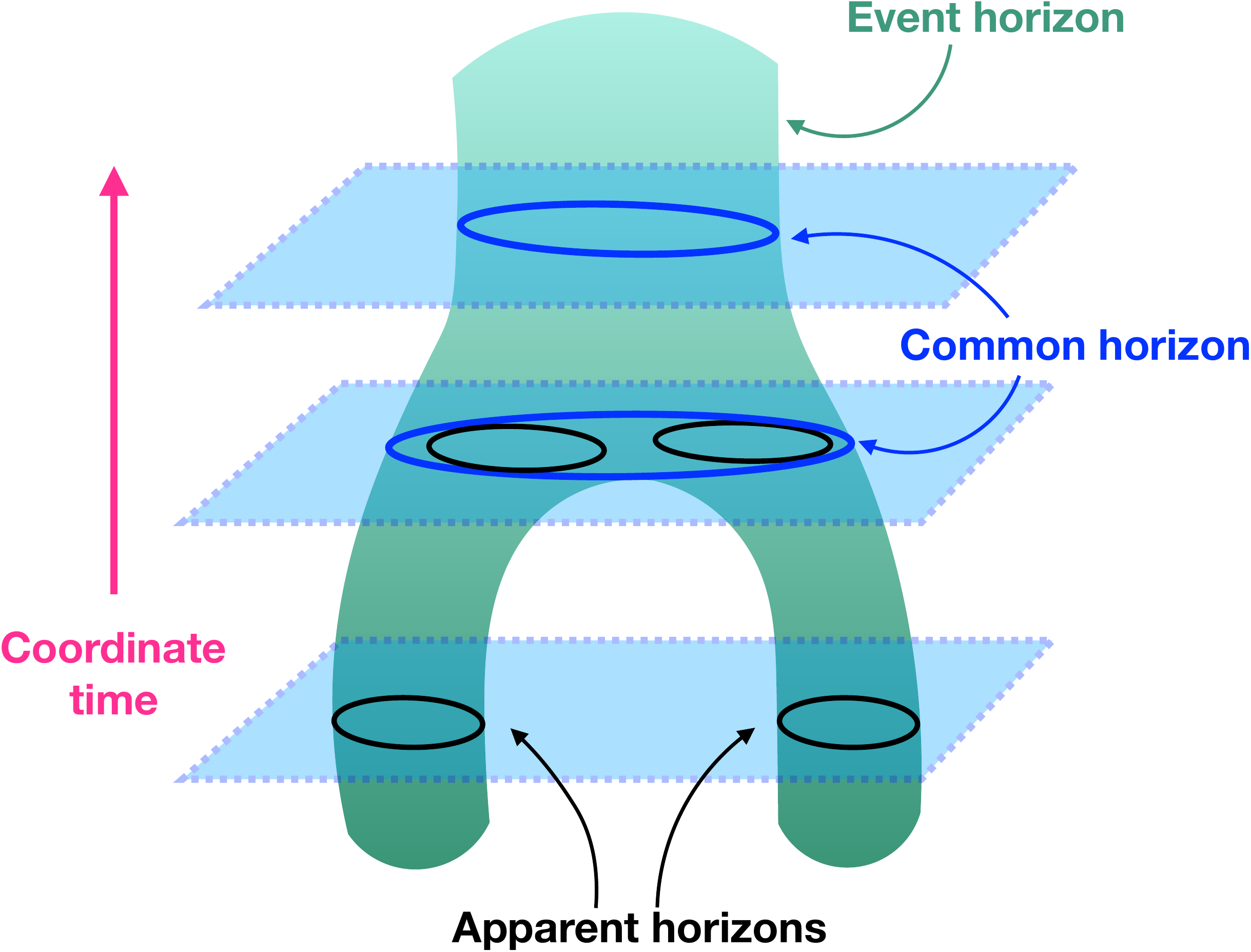}
  \caption{Schematic of horizons during a binary black hole merger, as detailed in Sec.~\ref{sec:horizons}. Each of the light blue planes corresponds to a 3-dimensional spatial slice of the simulation (with one dimension suppressed), and the coordinate time of the simulation is increasing from the bottom of the figure to the top. The green, `pants-like' region corresponds to the event horizon, which is a 3-dimensional hypersurface. Anything that enters the event horizon, including non-linearities, will not make it out to future null infinity, where the gravitational wave detectors are located. Each leg of the pants corresponds to each black hole, while the top of the pants corresponds to the final black hole after merger. On each slice, we show the 2-dimensional apparent horizons, which are always located inside of the event horizon. Hence, any non-linearities that enter the apparent horizons will not reach the gravitational wave detector. Initially, each black hole has an apparent horizon (blue circles), and during merger, a common horizon (blue circle) forms, encompassing both of the individual horizons. In numerical simulations, the region inside of the outermost apparent horizon on the slice is excluded from the computational domain. 
  }
  \label{fig:Horizons}
\end{figure}

Since we are interested in where non-linearities `go', it is worth taking some time to discuss black hole horizons. We provide a cartoon illustration of event and apparent horizons in Fig.~\ref{fig:Horizons}. 

Recall that the event horizon in a 4-dimensional spacetime is a 3-dimensional hypersurface that separates the events that can emit null rays that propagate to future null infinity from those that cannot (cf.~\cite{Wald:106274}). If gravitational radiation enters a black hole event horizon, it cannot then escape to future null infinity, where gravitational wave detectors live. Thus, if we look at whether non-linearities in a binary black hole simulation are inside or outside of the event horizon, we can see whether they will make it out to the gravitational wave detector. For a binary black hole simulation, the event horizon looks like a `pair of pants' (cf.~\cite{Cohen:2008wa, Bohn:2016soe}), where each pant leg corresponds to one of the two black holes, and the two legs combine smoothly to form the final black hole. 

In order to find the event horizon in a numerical relativity simulation, however, we need access to the entire history of the spacetime (recall that the event horizon is a 3-dimensional hypersurface). While the event horizon can be found during post-processing of a simulation (cf.~\cite{Bohn:2016afc, Bohn:2016soe, Cohen:2008wa}), an easier computation to perform during a simulation is to consider the \textit{apparent horizon}. The apparent horizon is a 2-dimensional surface of zero expansion on each slice of a black hole spacetime (cf.~\cite{baumgarteShapiroBook}). The apparent horizon is relatively simple to find on each slice of a numerical relativity simulation~\cite{Thornburg:1995cp, Gundlach:1997us, Lovelace:2014twa}. A key aspect of apparent horizons is that they always lie \textit{inside} of or coincide with the event horizon. Thus, if some non-linearities enter the apparent horizon, we know that they have entered the event horizon as well. 

In binary black hole spacetime there are initially two apparent horizons on early slices, one for each black hole. As the black holes merge, a \textit{common} apparent horizon, which encompasses both of the individual black hole horizons, forms. The precise coordinate time at which this occurs in slicing-dependent (cf.~\cite{Bohn:2016soe}), and hence the time of formation of the common horizon is not physically meaningful. However, the common horizon lies inside of the event horizon, which is a physically meaningful surface. Thus, any non-linearities that enter the common horizon have entered the even horizon as well, and will not make it out to the gravitational wave detector. 

Finally, we must make a computational aside on horizons, having to do with \textit{excision regions}. In the code that we use in this study, the Spectral Einstein Code (SpEC)~\cite{SpECwebsite}, we \textit{excise} the black hole singularities from the computational domain by excluding the regions inside of the apparent horizons from our computational grid. Because the excised region is causally disconnected from the exterior, this does not affect the data on the computational domain. Once the common horizon forms, everything inside of the common horizon is excised from the computational domain. The actual procedure is more subtle, requiring making sure that the excision boundaries are surfaces with no ingoing characteristics into the computational domain. Technical details are given in~\cite{Hemberger:2012jz}, but for the purposes of our study, we note the excision surfaces are always inside or coincide with the apparent horizons, and thus any non-linearities that leave the computational domain by way of the excision region are inside of the apparent horizons, which in turn are inside of the event horizon. Thus when we visualize the Kerrness measures in the strong-field region of a binary black hole merger, we will show the excision region, and anything that enters the excision region has entered the event horizon.

\section{Results}
\label{sec:results}

\subsection{Simulation details}
\label{sec:simulation}

For this study, in order to isolate the behavior of non-linearities, we consider the simplest system: an equal mass, non-spinning, circular binary, choosing an initial orbital frequency of $\Omega_0 = 0.025/M$. We use the Spectral Einstein Code~\cite{SpECwebsite}, with the methods given in~\cite{SXSCatalog, Lindblom2006, Scheel:2008rj, Szilagyi:2009qz, Hemberger:2012jz} to simulate the binary. We perform simulations for low, medium, and high resolutions, with each resolution $n$ decreasing the truncation error tolerance by a factor of $4^{-n}$ (cf.~\cite{SXSCatalog} for more details). We check that the behavior presented in this study is convergent with numerical resolution. The common apparent horizon (cf. Sec.~\ref{sec:horizons}) forms at coordinate time $t = 1127.0\,M$, and the final black hole has a Christodolou mass $0.9517\,M$ and a final dimensionless spin of $0.686$.

\subsection{Linearities in the wave-zone}
\label{sec:overtones}

\begin{figure}
  \includegraphics[width=\columnwidth]{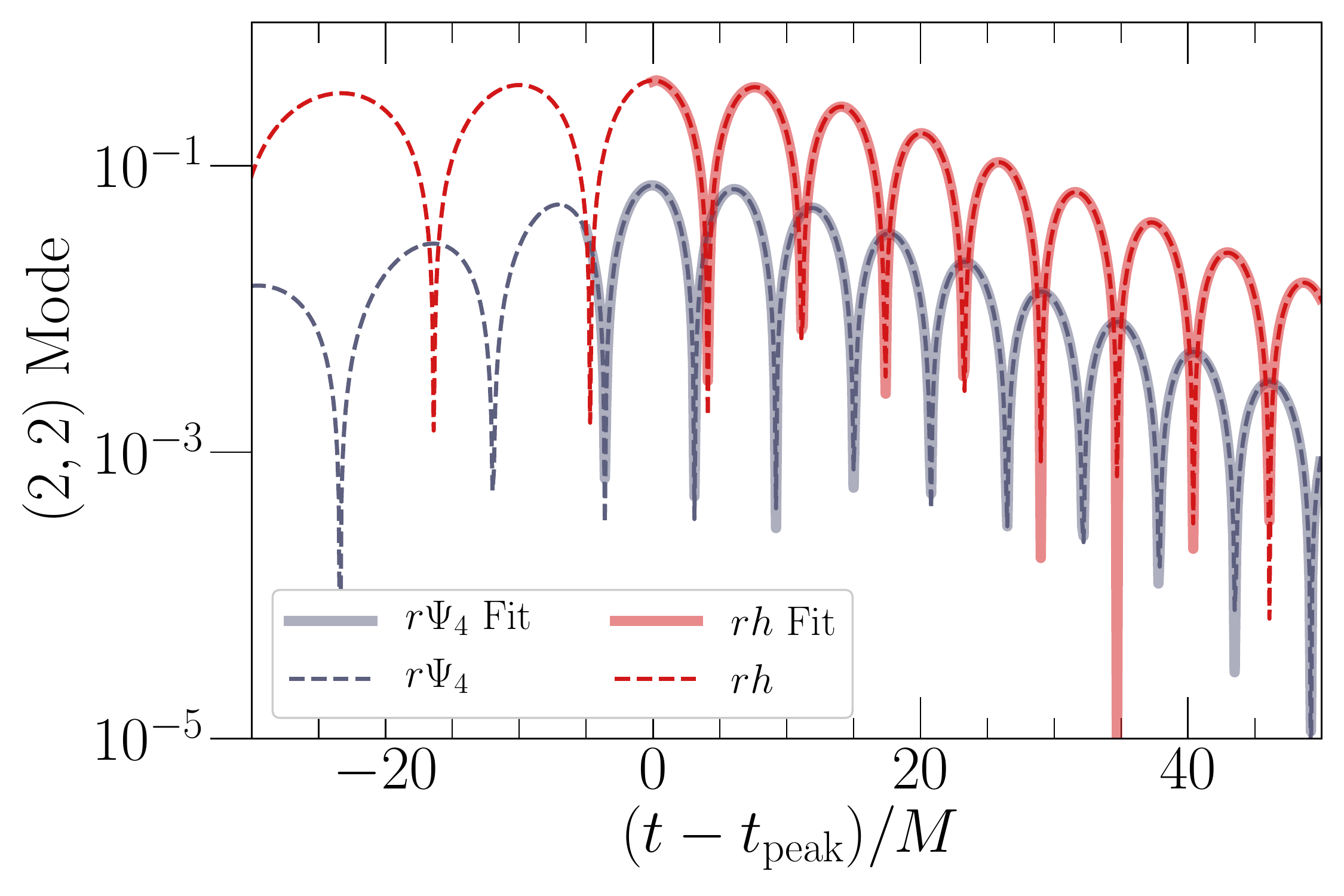}
  \caption{
    Quasi-normal mode fits to the post-merger gravitational radiation, as described in Sec.~\ref{sec:overtones}. We show the dominant $(2,2)$ mode of the extrapolated Newman-Penrose scalar $r \Psi_4$ (dashed gray) and the strain $rh$ (dashed red) as a function of time to the peak of $rh$. We fit a combination of overtones of the $(2,2)$ mode in each case (solid lines), showing that the fit is faithful all the way back to the peak of $rh$. In this figure, we use a combination of 4 overtones, but we check that the fit improves with the addition of more overtones. 
  }
  \label{fig:Overtones}
\end{figure}

The post-merger gravitational waveform for this simulation can be fit by linear theory, using a sum of quasi-normal modes including overtones all the way back to the peak of the waveform. In Fig.~\ref{fig:Overtones}, we show fits for the dominant $(2,2)$ mode of the Newman Penrose scalar $r\Psi_4$ and the gravitational wave strain $rh$ using 4 overtones, both extrapolated to infinity using the methods in~\cite{Boyle:2009vi}. We use the fitting methods detailed in~\cite{MashaHeadOn}, with the quasi-normal mode frequencies for each overtone given by~\cite{QNMCode}. This demonstrates that far from the black holes, the gravitational radiation is fully described by linear theory.

\subsection{Non-linearities in the strong-field region}
\label{sec:nonlinear}

Given that the post-merger gravitational radiation is well-described by linear theory, let us now investigate the presence of non-linearities in the strong-field region. As outlined in Sec.~\ref{sec:methods}, each of the Kerrness measures considered in this paper can be used to measure non-linearities in the spacetime. For each quantity, we will present its value on a set of spatial slices during the merger, with each slice being labeled by a coordinate time of the simulation. While our simulations are fully 3-dimensional, for ease of visualization, we present the quantities in the plane normal to the orbital angular momentum of the binary (note that there is no precession in our non-spinning simulation). We will present three of the Kerrness measures given in~\cite{Bhagwat:2017tkm} (due to the fact that the others require taking higher numerical derviatives and hence are more prone to numerical noise). 

As explained in Sec.~\ref{sec:horizons}, we excise the portion of the spacetime inside of the apparent horizons and common horizon from our simulations. This region thus does not show up in the visualizations. Most importantly, this excision region lies \textit{inside} of the event horizon, so anything that enters this region will not make it out to future null infinity. Thus, any non-linearities which enter this region will not make it to the gravitational wave detector. 

We are most-interested in the \textit{relative} values of the Kerrness measures as a function of time. Recall from Sec.~\ref{sec:numerical_details} that the lower bound of the scale is given by the numerical noise floor, and can be thought of as zero. We can thus compare the maximum values of each Kerrness measure to this `zero' value to quantify how many orders of non-linearity are spanned during the merger.

\begin{figure*}
\subfloat{\includegraphics[width=0.24\textwidth]{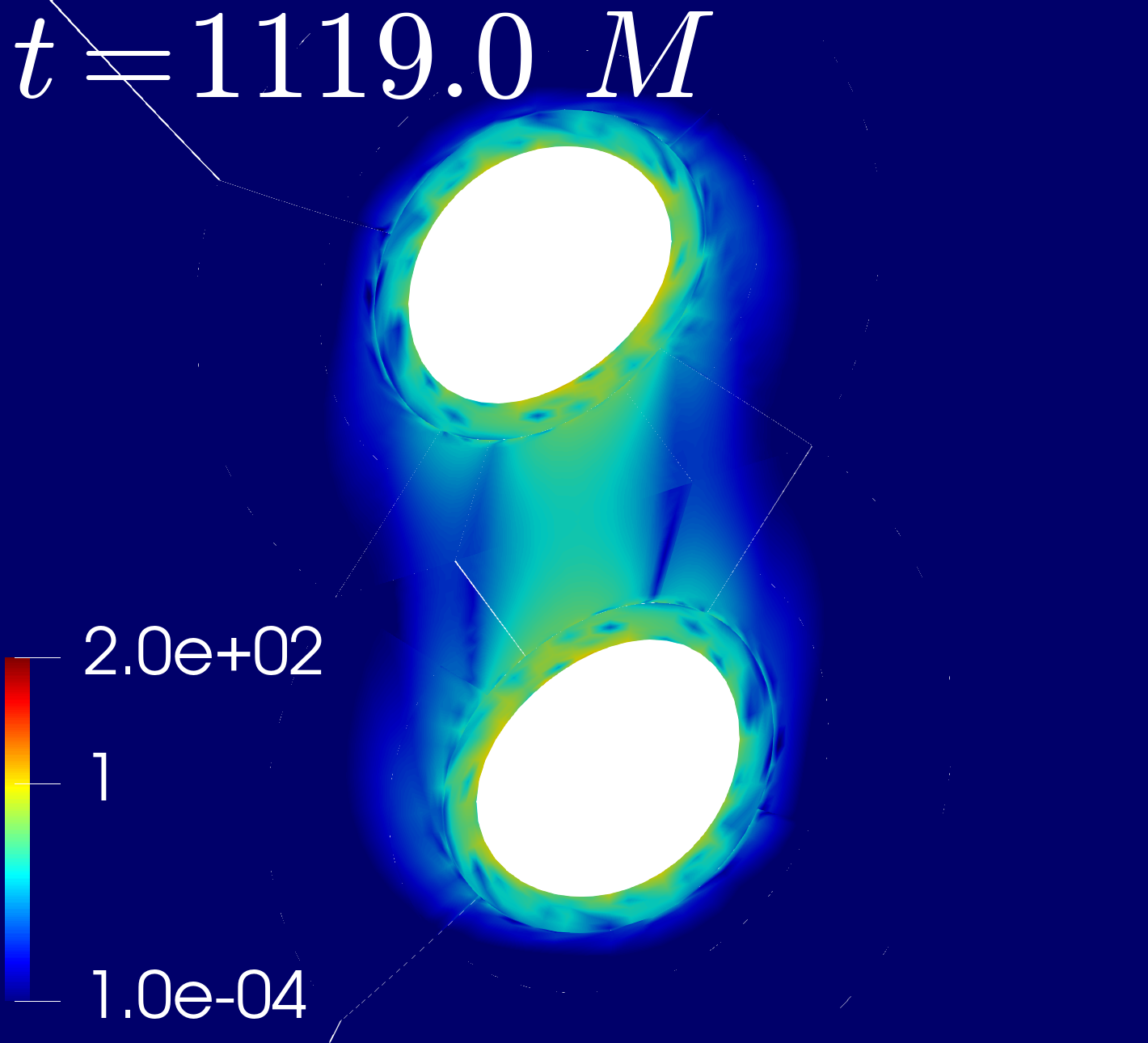} } \hspace{-0.15cm}
\subfloat{\includegraphics[width=0.24\textwidth]{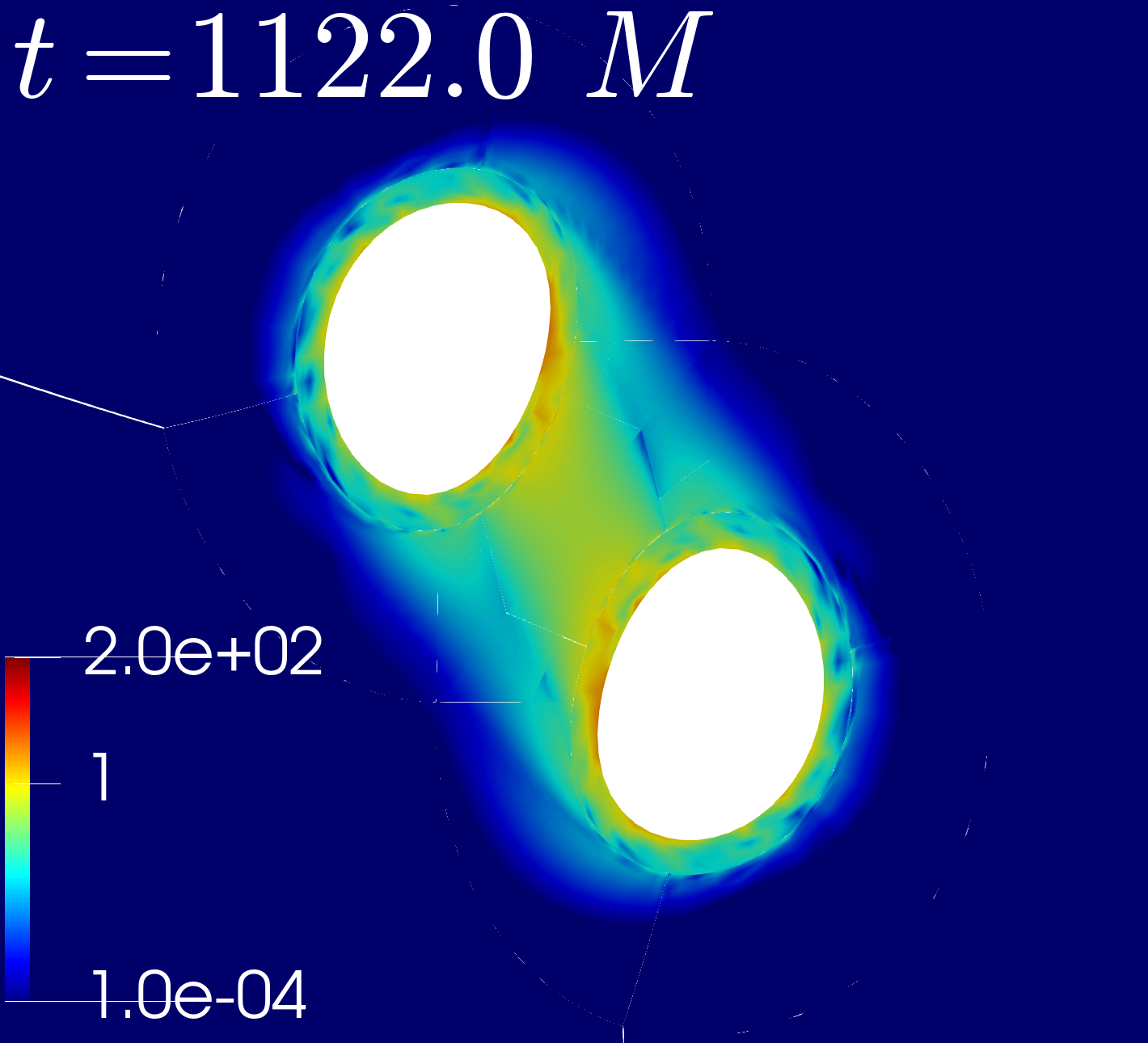} }  \hspace{-0.15cm}
\subfloat{\includegraphics[width=0.24\textwidth]{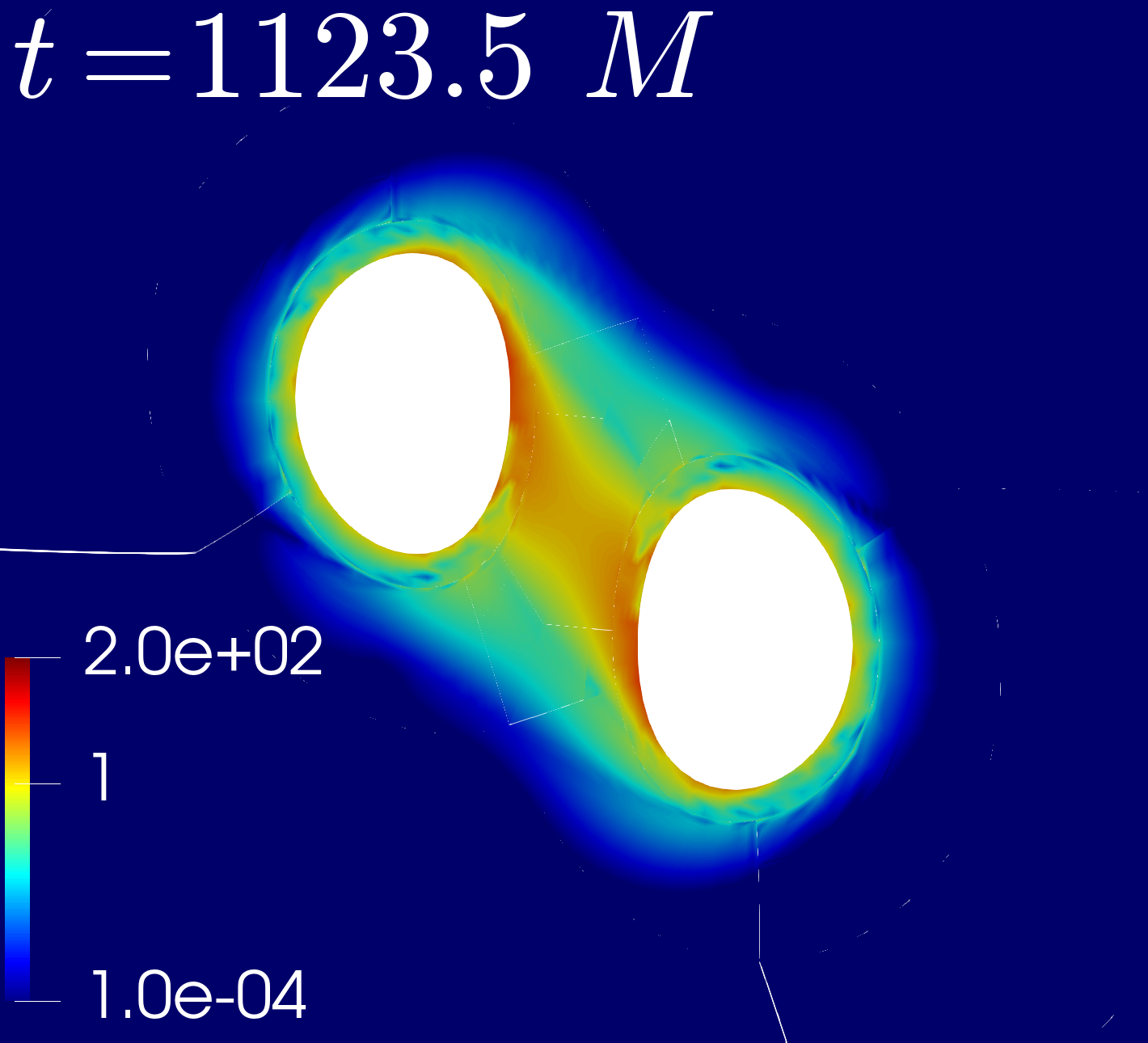} } \hspace{-0.15cm}
\subfloat{\includegraphics[width=0.24\textwidth]{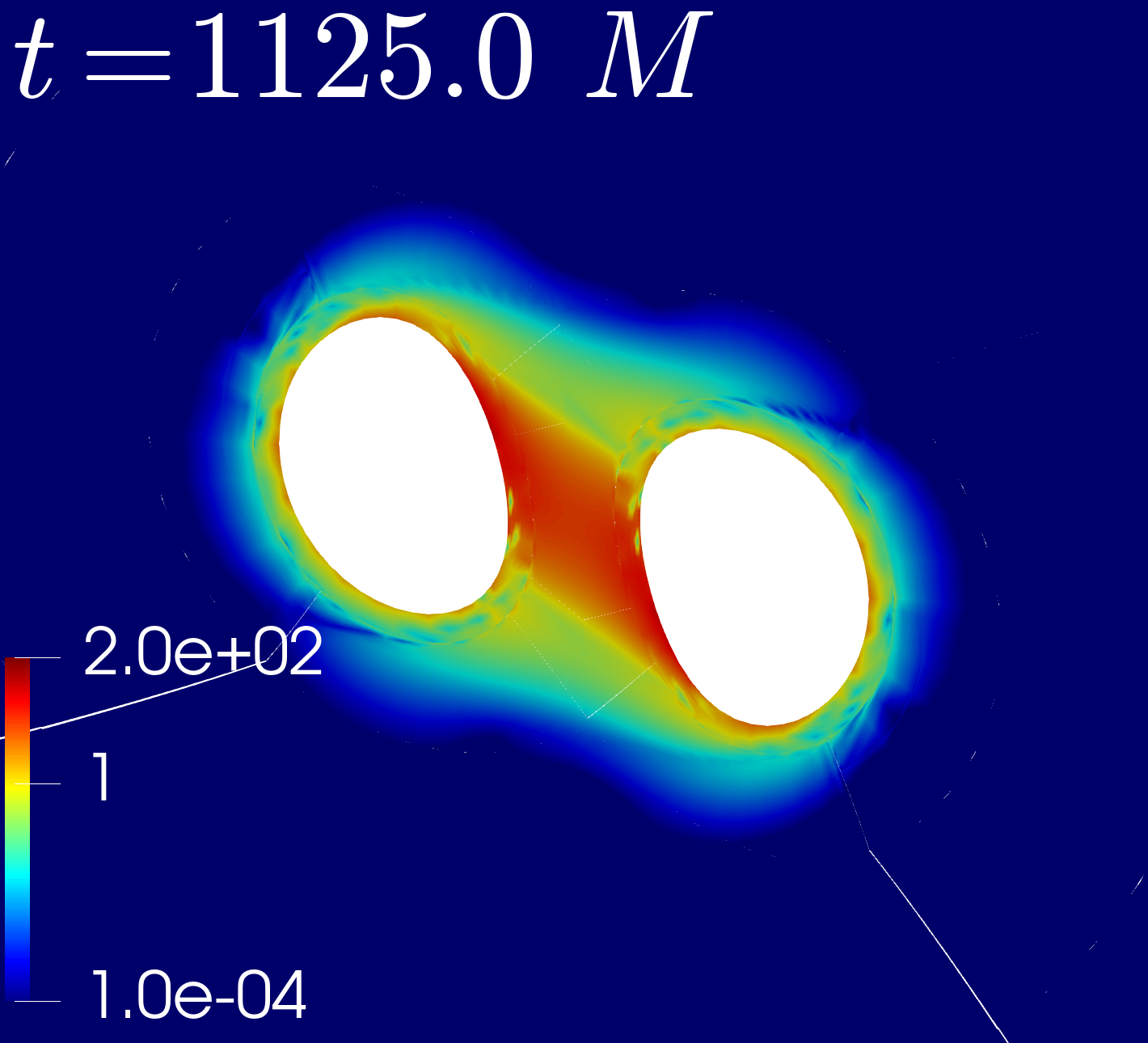} } \\ [-0.35cm]
\subfloat{\includegraphics[width=0.24\textwidth]{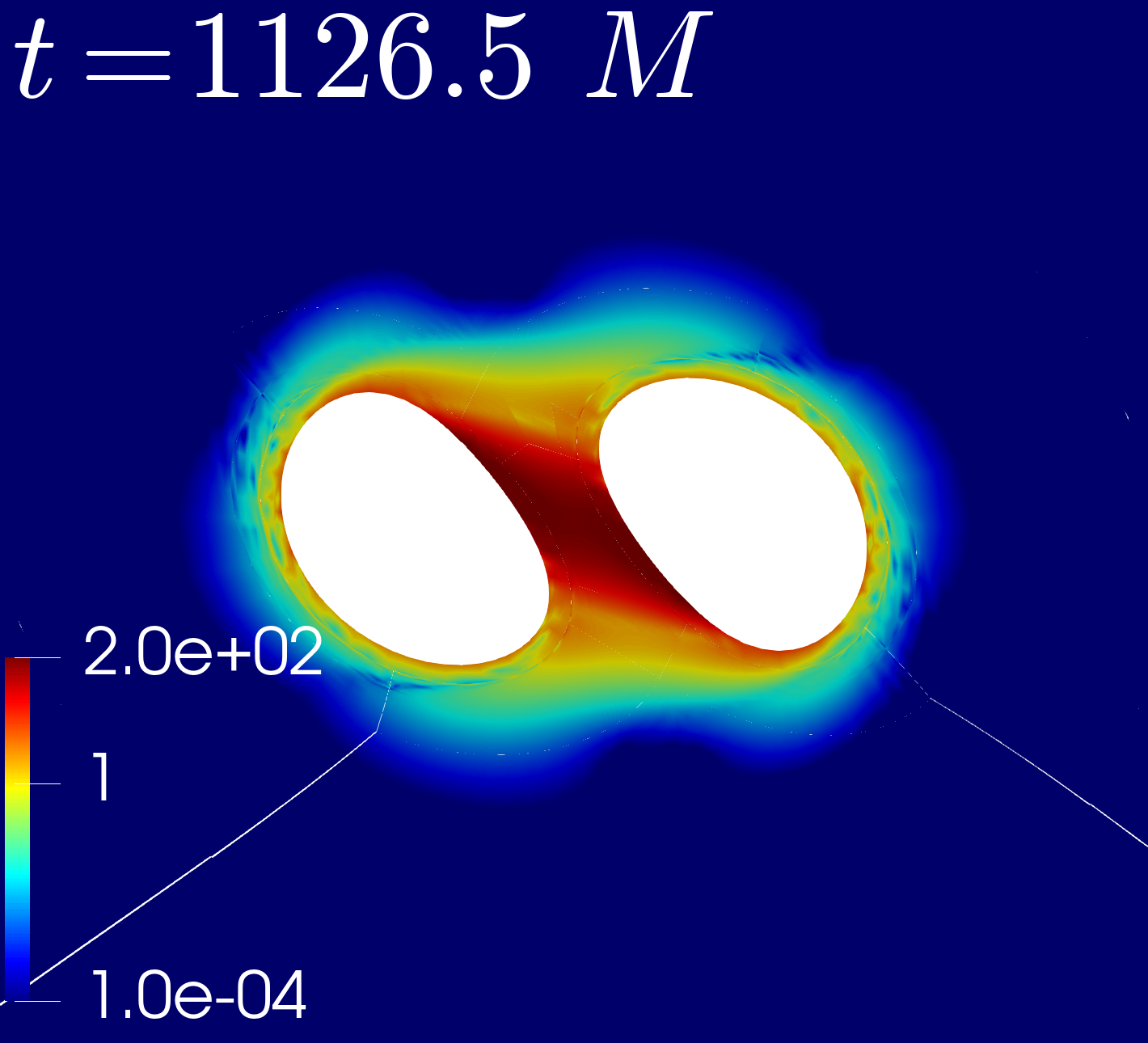} } \hspace{-0.15cm}
\subfloat{\includegraphics[width=0.24\textwidth]{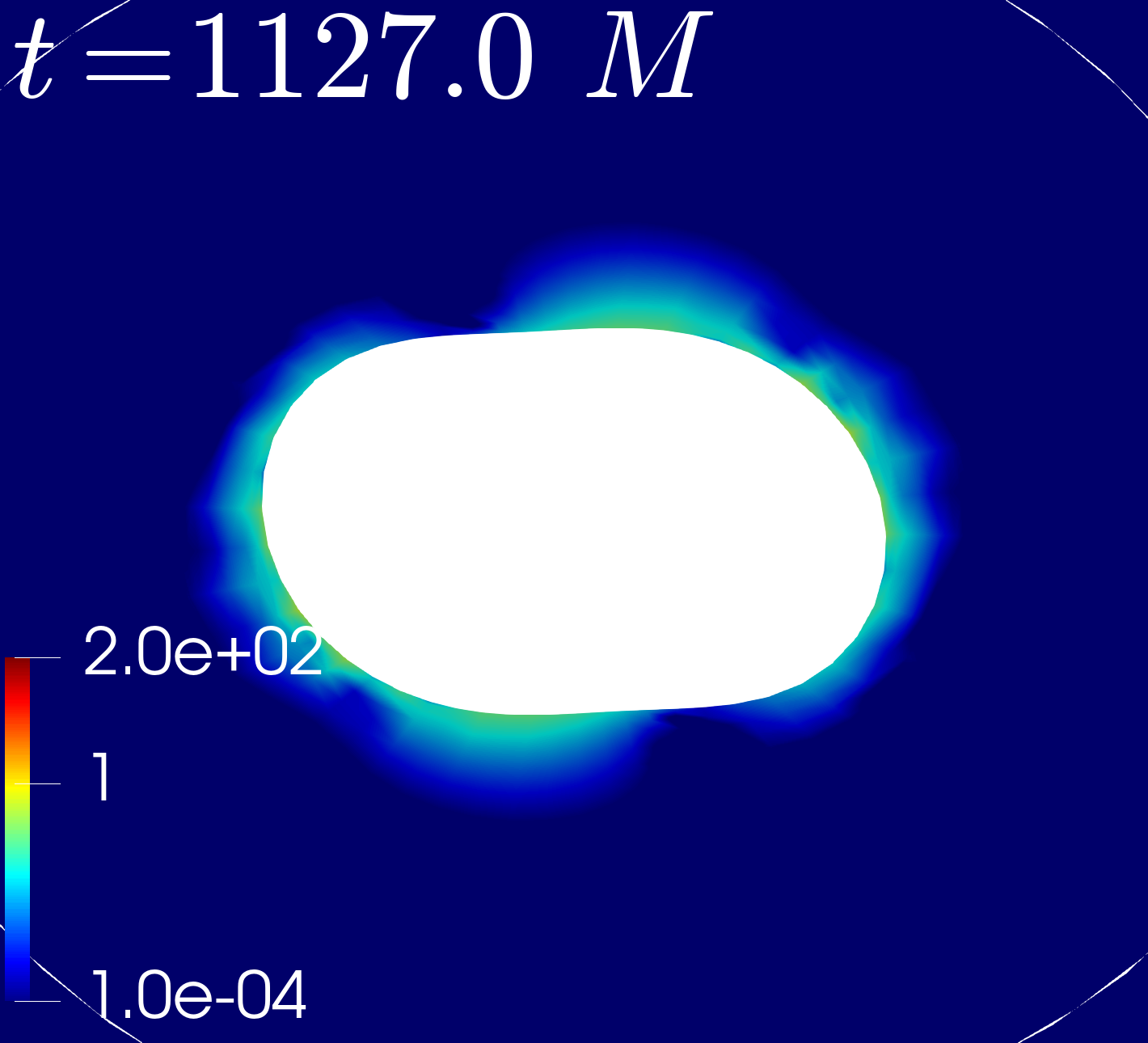} }  \hspace{-0.15cm}
\subfloat{\includegraphics[width=0.24\textwidth]{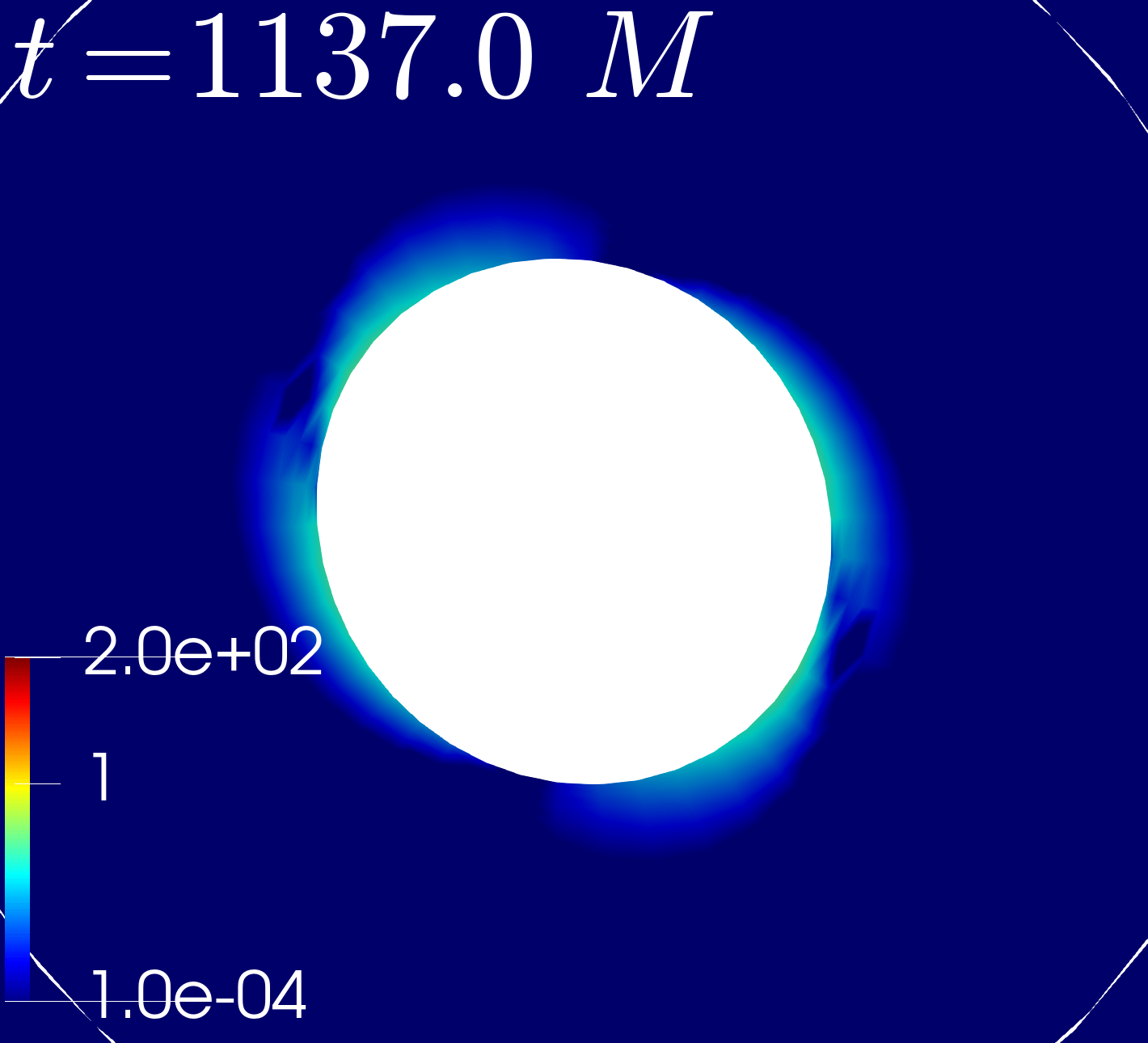} } \hspace{-0.15cm}
\subfloat{\includegraphics[width=0.24\textwidth]{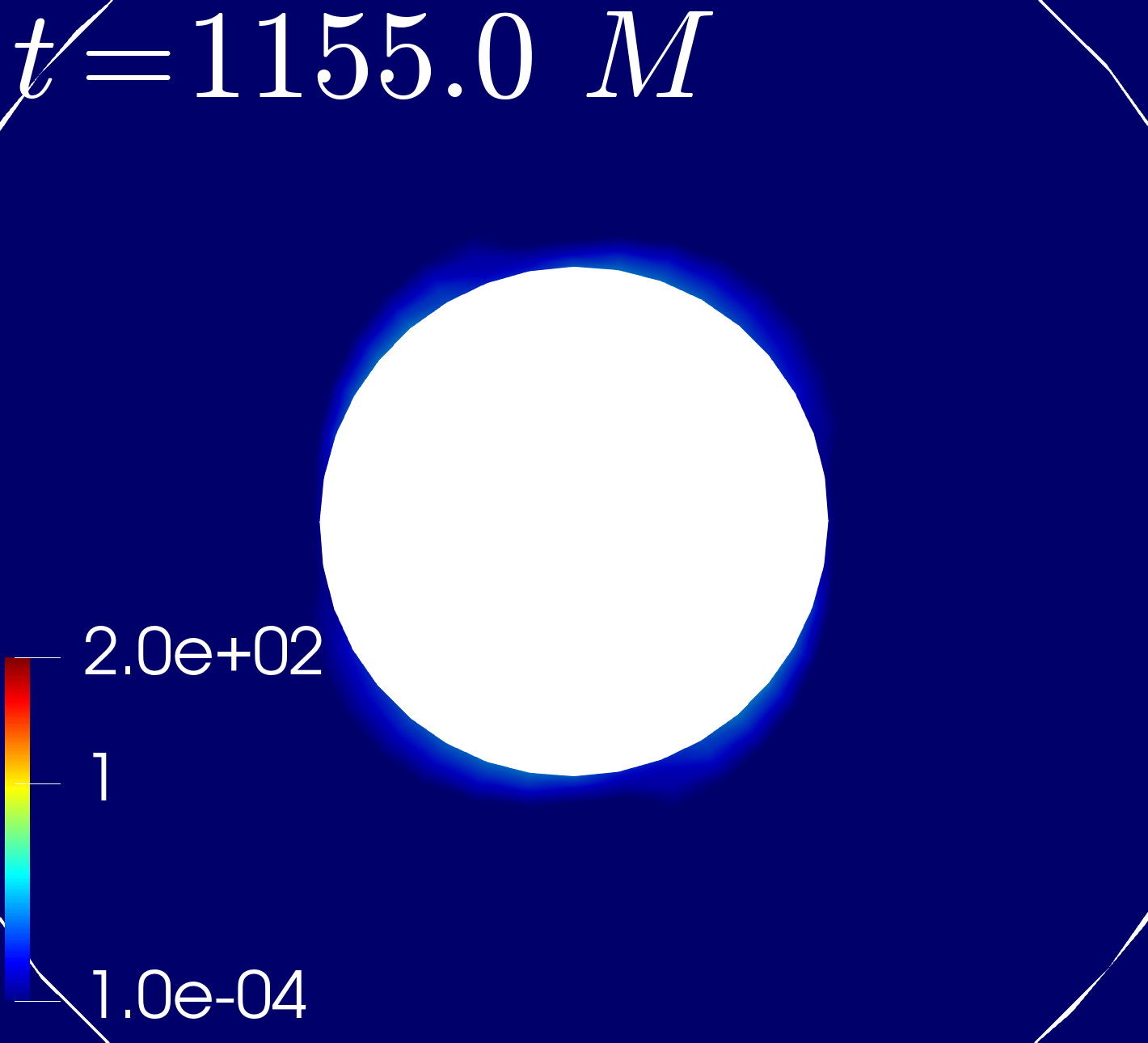} } 
\caption{Visualization of non-linearities using the \textbf{Speciality} Kerrness measure from~\cite{Bhagwat:2017tkm} on the computational domain of an equal mass, non-spinning, circular binary black hole merger. Increasing values correspond to higher non-linearity, but should be thought of as a relative scale. Each panel corresponds to a different spatial slice labelled by coordinate time $t$ of the simulation. We show the Kerrness measure in the plane of the binary normal to the orbital angular momentum. The white regions correspond to the computational domain excision regions, as detailed in Sec.~\ref{sec:horizons}. These are located inside of the black hole apparent horizons, which are in turn located inside of the event horizon. Thus, anything that enters this region will not make it out to the gravitational wave detector. Initially, there are two excision regions, one corresponding to each black hole. Between $t = 1126.5\,M$ and $t=1127.\,M$, a \textit{common} apparent horizon forms (cf. Fig.~\ref{fig:Horizons}), encompassing both black holes. We thus excise the region inside of the common horizon from the computational domain, noting that this common horizon is inside of the event horizon, and hence anything that enters this region will not make it out to the gravitational wave detectors. Note that the colors are on a logarithmic scale. The lower bound of the color scale (dark blue) is set by the numerical noise floor (cf. Sec.~\ref{sec:numerical_details}). This corresponds to the fact that the Kerrness measures are non-zero on a pure Kerr black hole due to numerical error, but that they converge to zero with increasing numerical resolution. Thus, at each numerical resolution, there is a `noise floor'. The values should be treated as relative, with an increase in six orders of magnitude visible in this case. We see that as the black holes merge, strong non-linearities develop, but that these non-linearities are mostly encompassed by the common horizon, and that any remaining ones enter the common horizon with time. At late times, a quiescent Kerr black hole remains. 
}
\label{fig:Speciality}
\end{figure*}

We present the first Kerrness measure in Fig.~\ref{fig:Speciality}. In the nomenclature of~\cite{Bhagwat:2017tkm}, this is the \textbf{Speciality} measure, which determines whether the spacetime is \textit{algebraically special} (cf.~\cite{stephani2009exact} for technical details). We see that initially, when the black holes are still relatively far apart, according to this measure, there are fewer non-linearities present. As the black holes come closer to merging, strong non-linearities develop between the two black holes. Once the common horizon forms, however, it encompasses most of these quantified non-linearities. As time progresses, the remaining non-linearities as quantified by the \textbf{Speciality} measure enter the common horizon, and hence enter event horizon and do not make it to future null infinity. At later times, we see a quiescent Kerr black hole. 

\begin{figure*}
\subfloat{\includegraphics[width=0.24\textwidth]{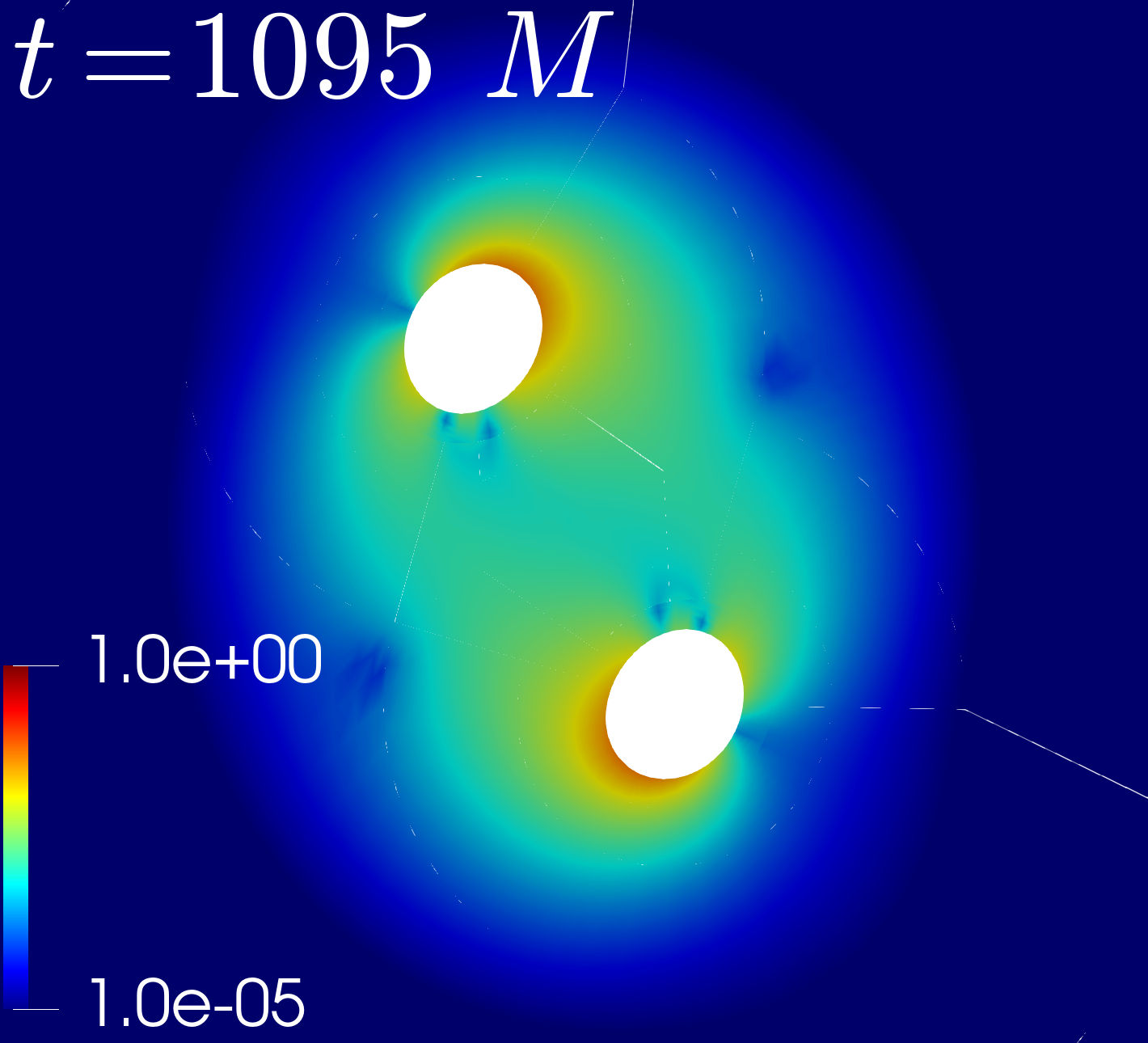} } \hspace{-0.15cm}
\subfloat{\includegraphics[width=0.24\textwidth]{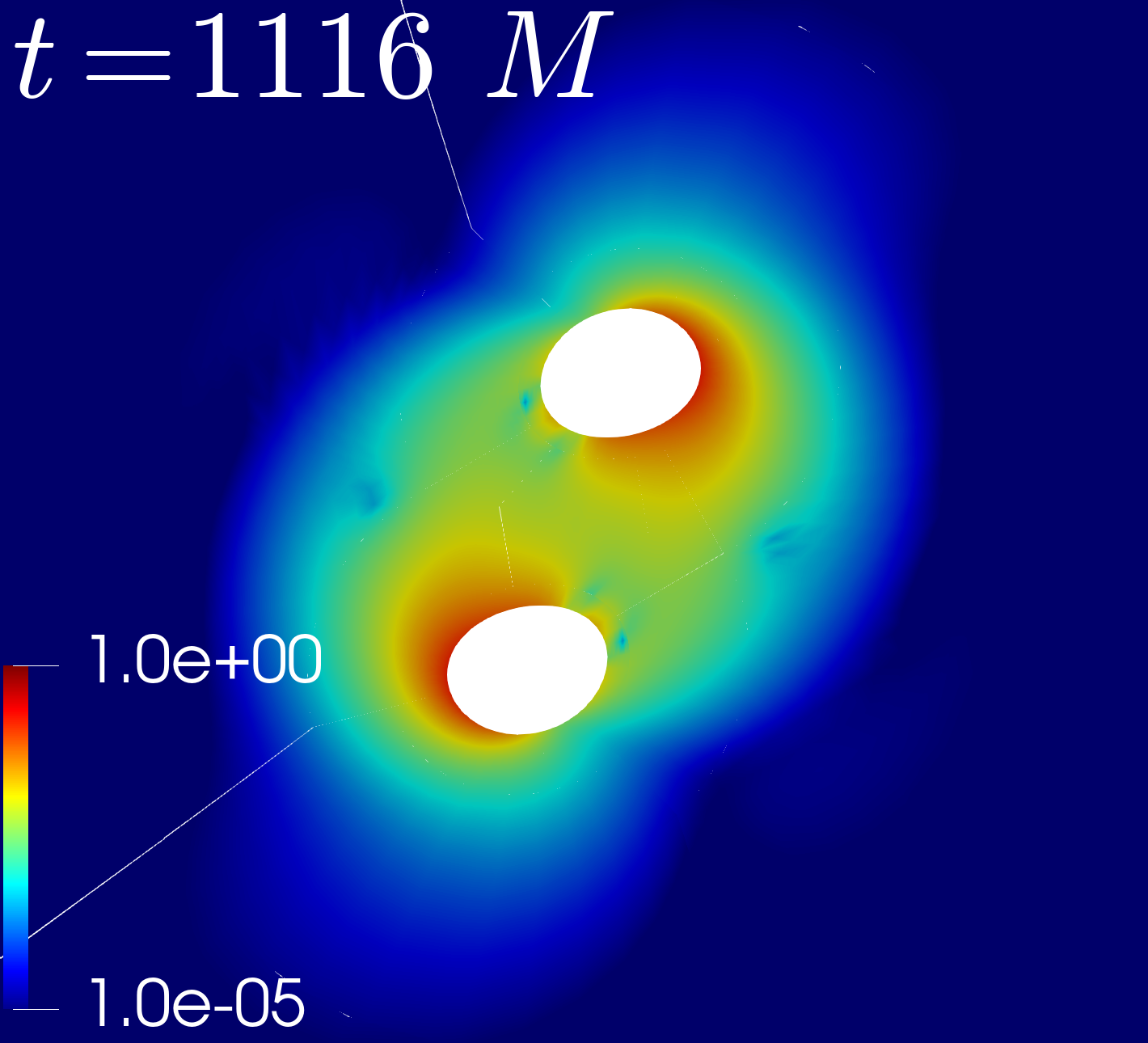} }  \hspace{-0.15cm}
\subfloat{\includegraphics[width=0.24\textwidth]{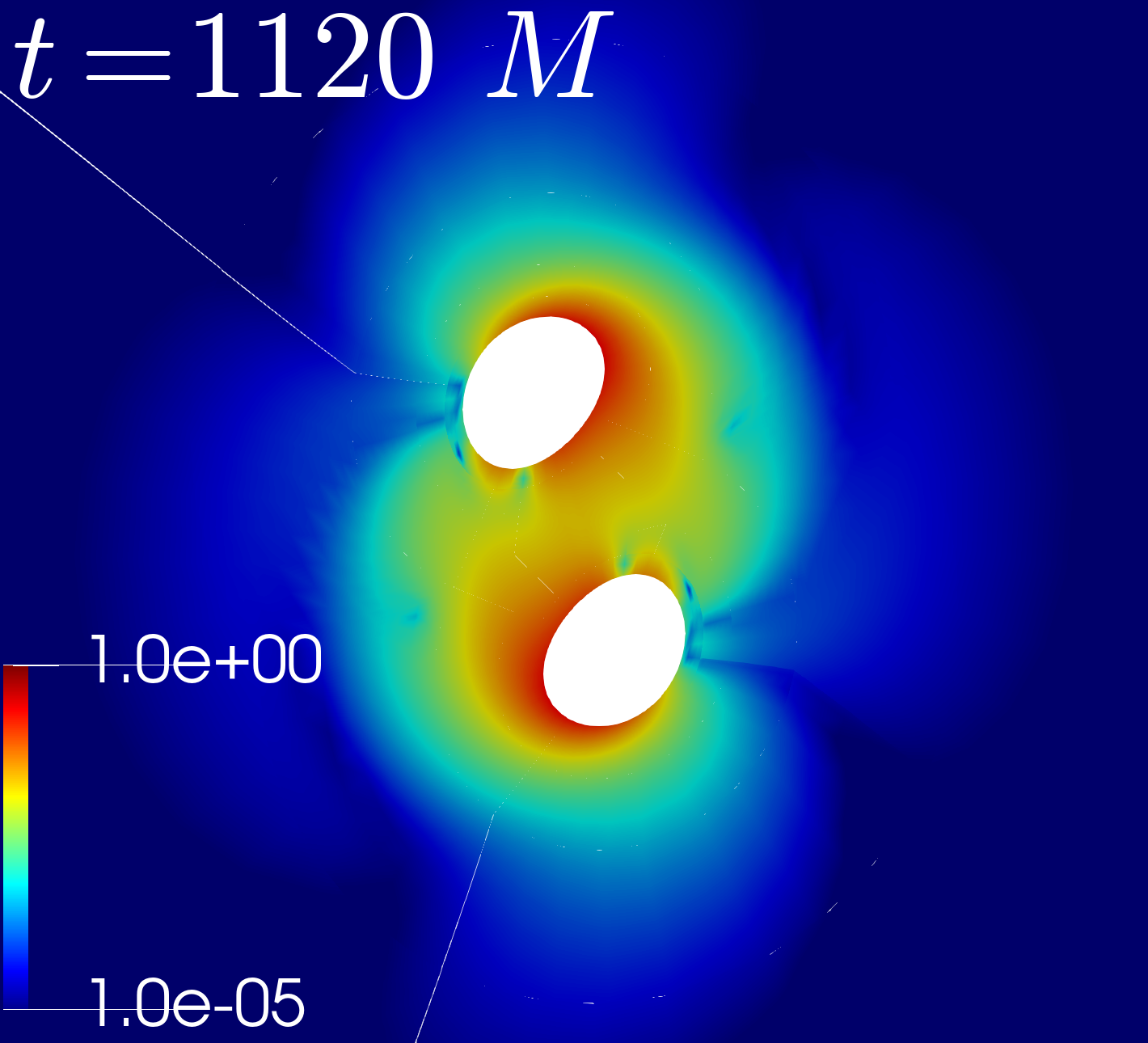} } \hspace{-0.15cm}
\subfloat{\includegraphics[width=0.24\textwidth]{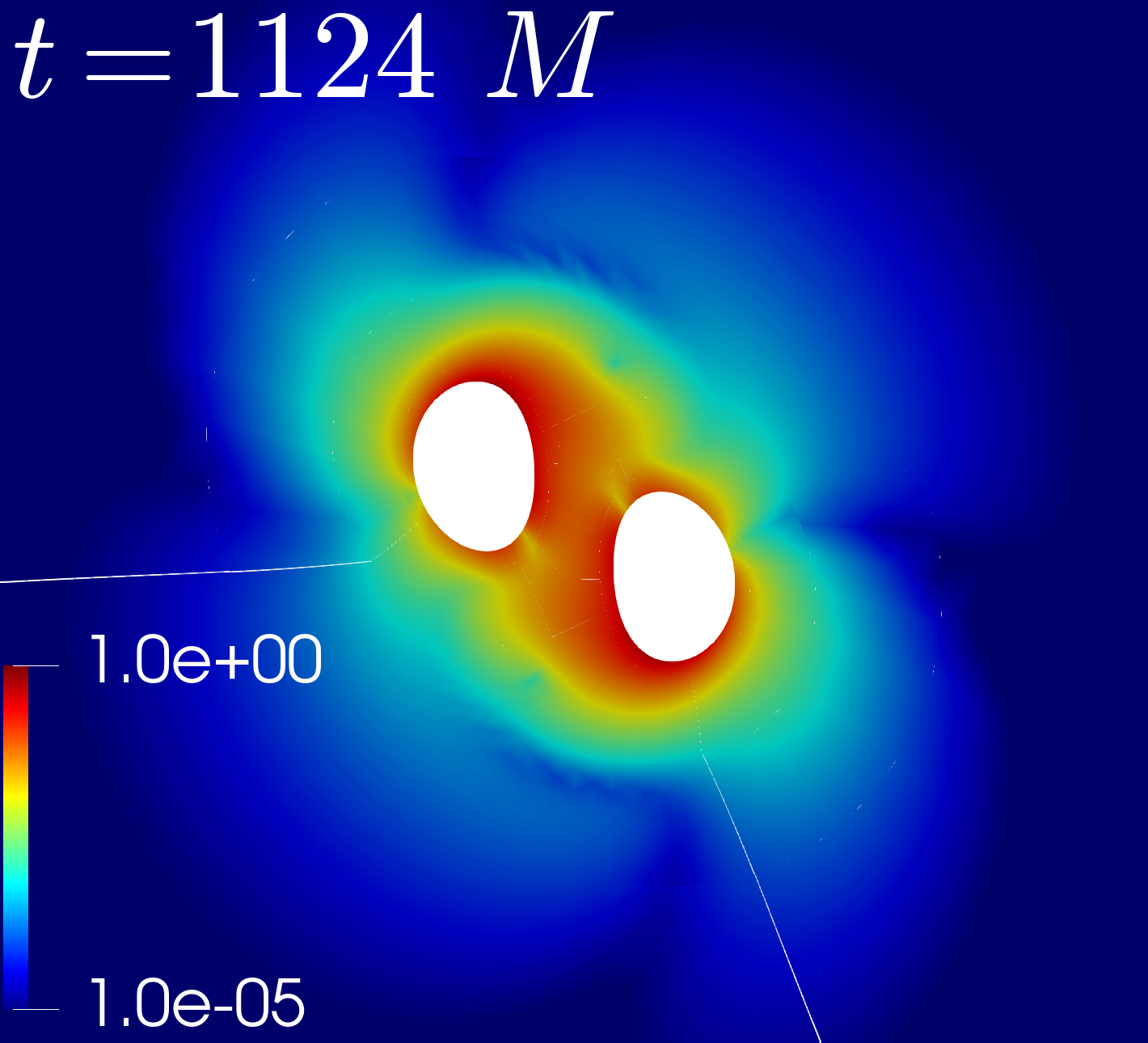} } \\ [-0.35cm]
\subfloat{\includegraphics[width=0.24\textwidth]{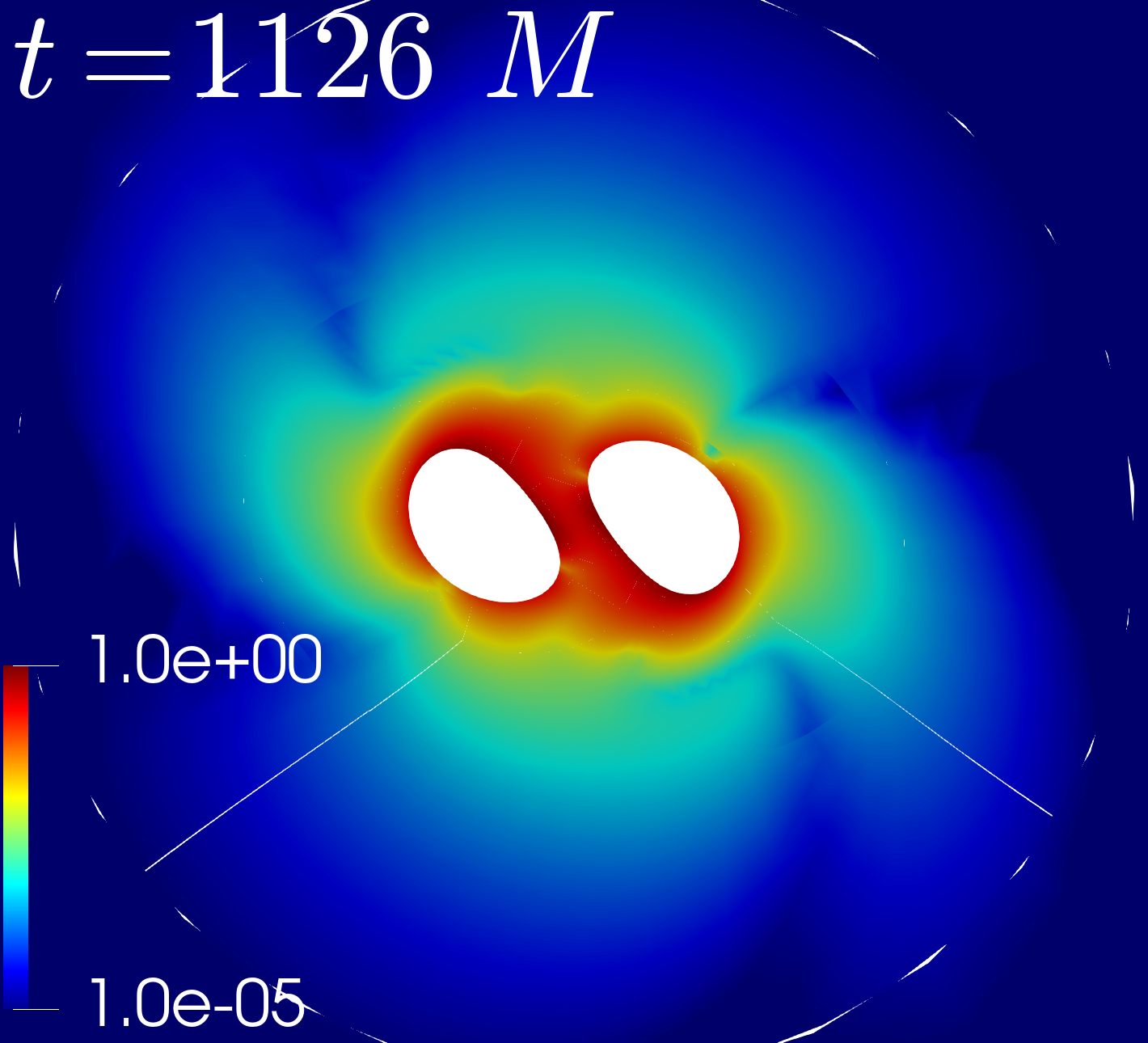} } \hspace{-0.15cm}
\subfloat{\includegraphics[width=0.24\textwidth]{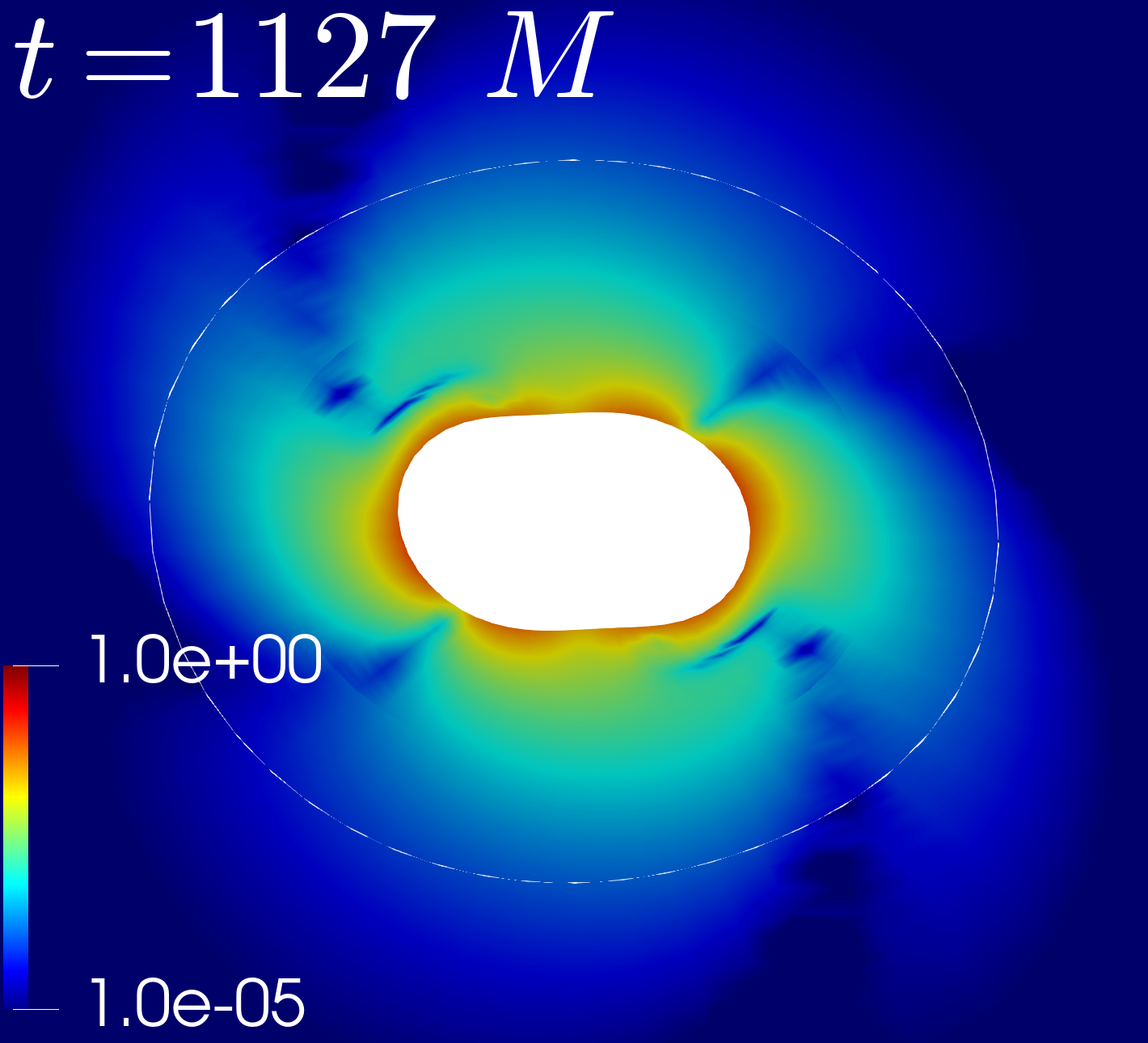} }  \hspace{-0.15cm}
\subfloat{\includegraphics[width=0.24\textwidth]{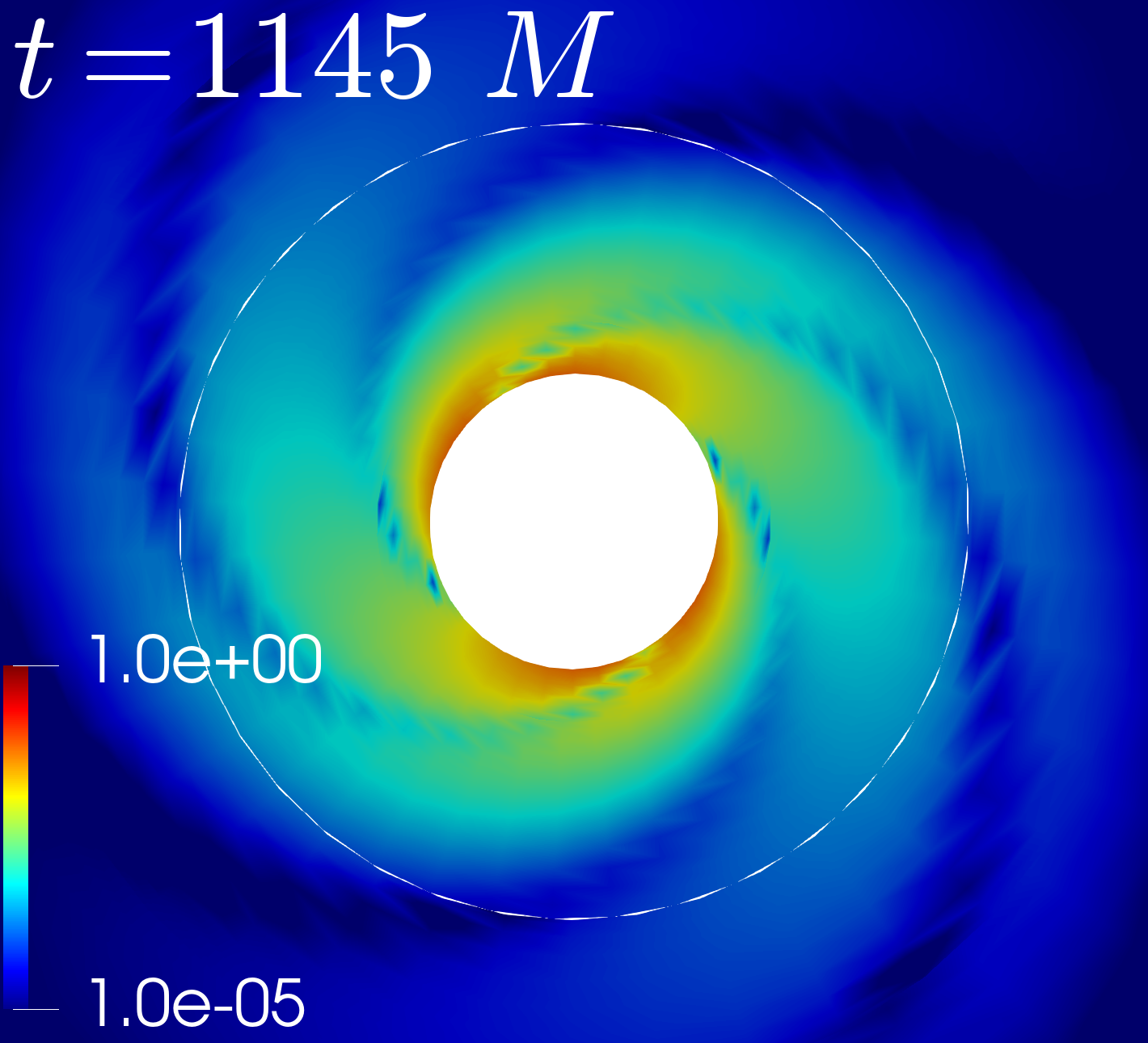} } \hspace{-0.15cm}
\subfloat{\includegraphics[width=0.24\textwidth]{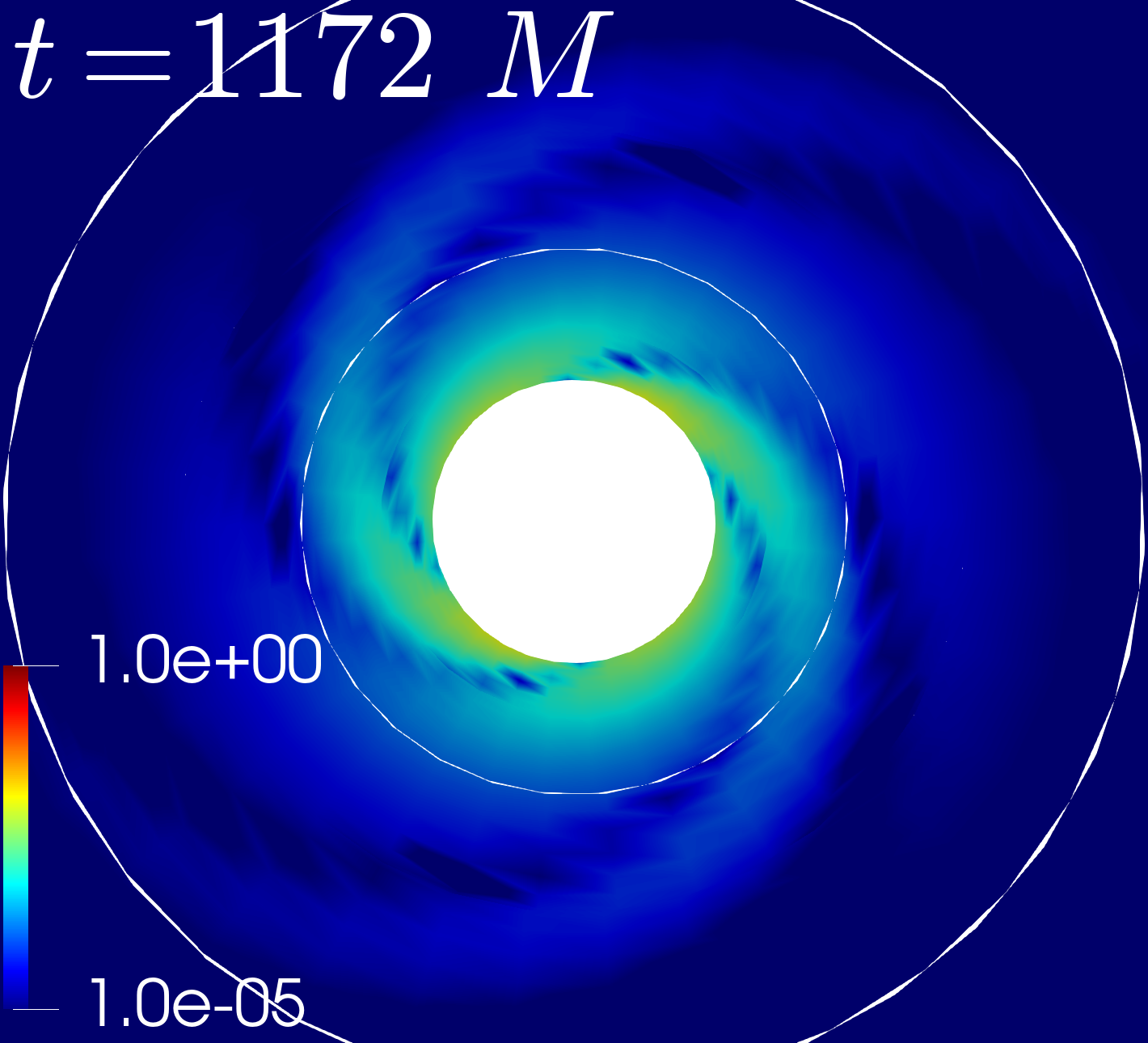} } \\ [-0.35cm]
\subfloat{\includegraphics[width=0.24\textwidth]{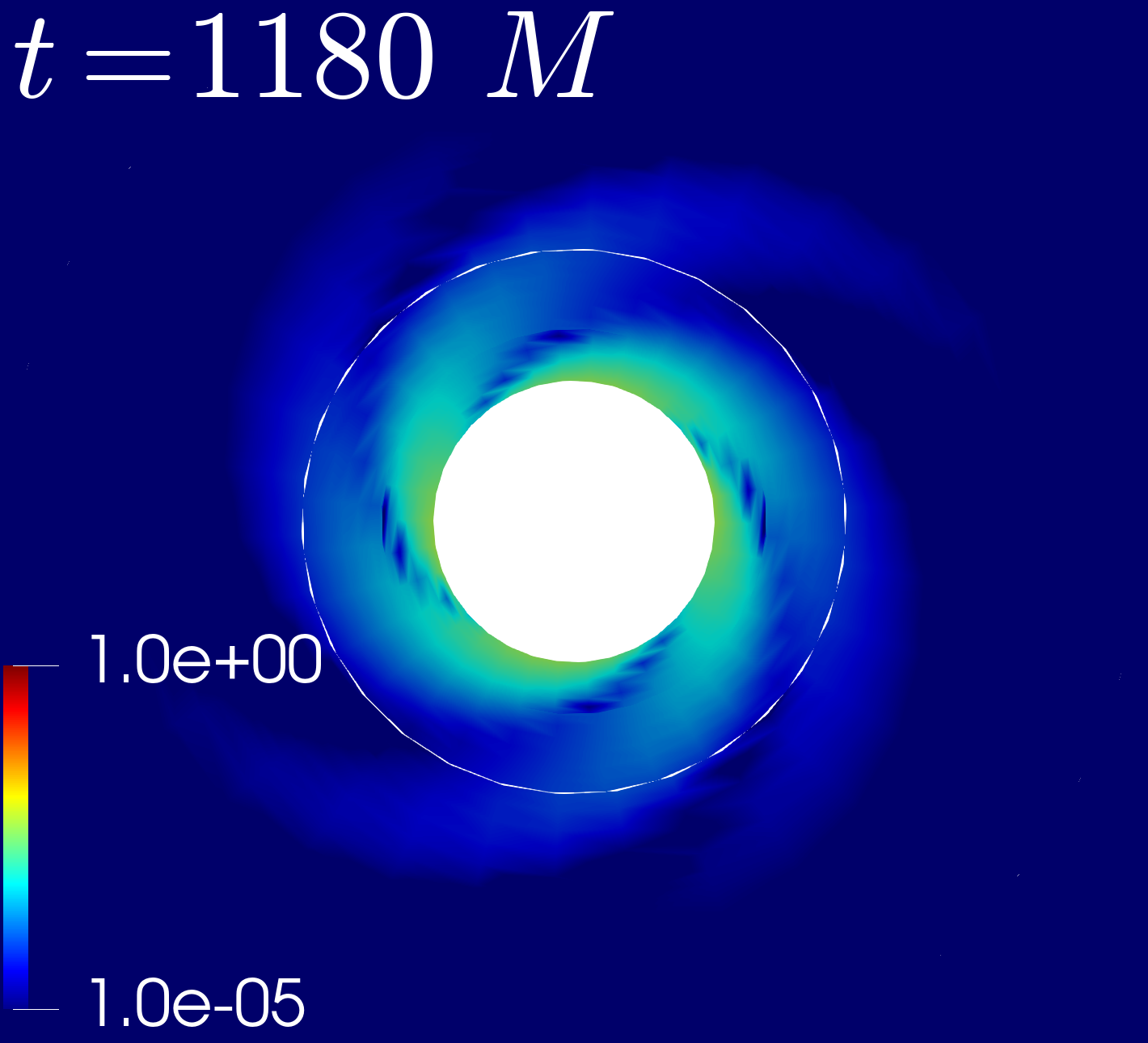} } \hspace{-0.15cm}
\subfloat{\includegraphics[width=0.24\textwidth]{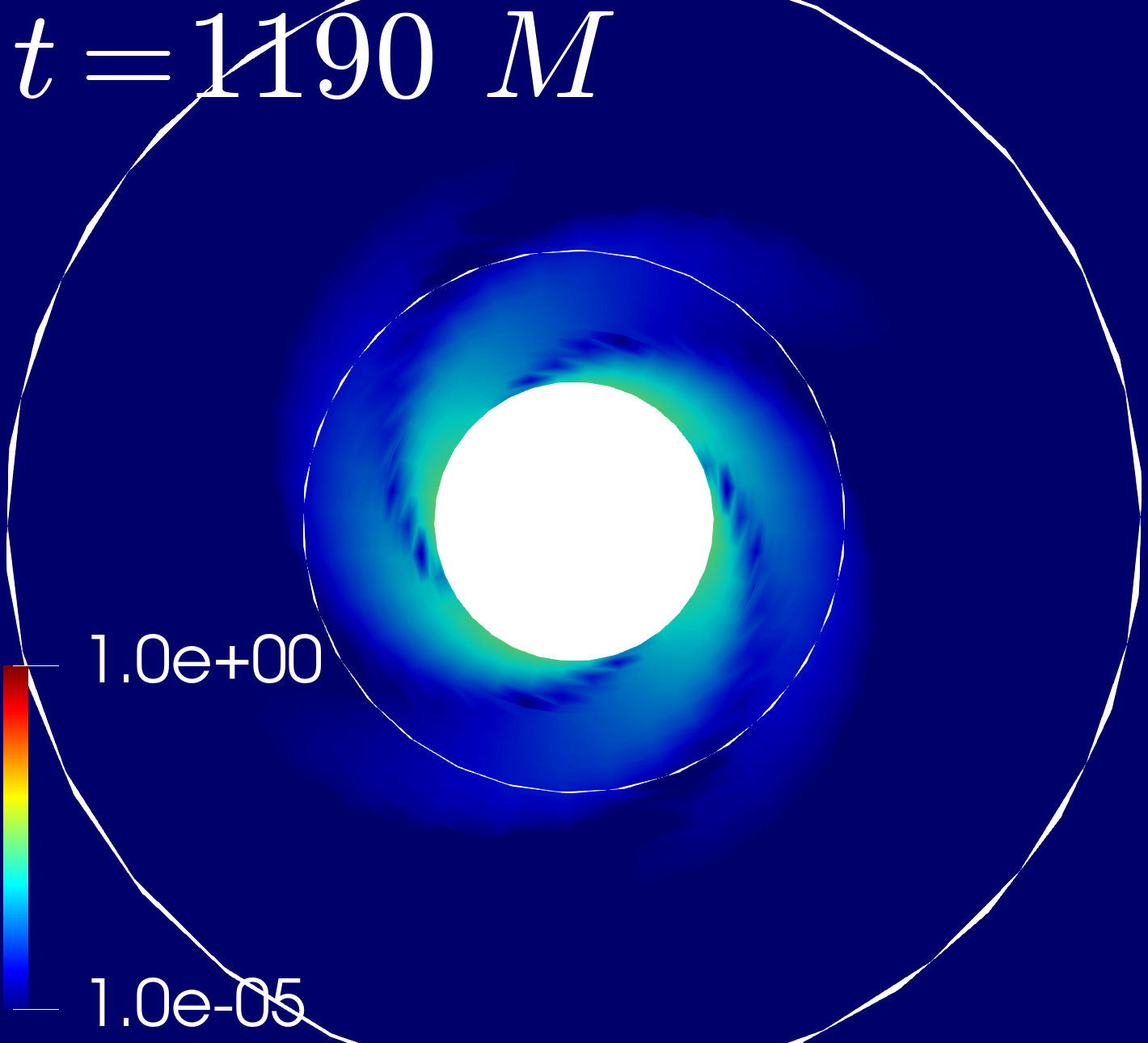} }  \hspace{-0.15cm}
\subfloat{\includegraphics[width=0.24\textwidth]{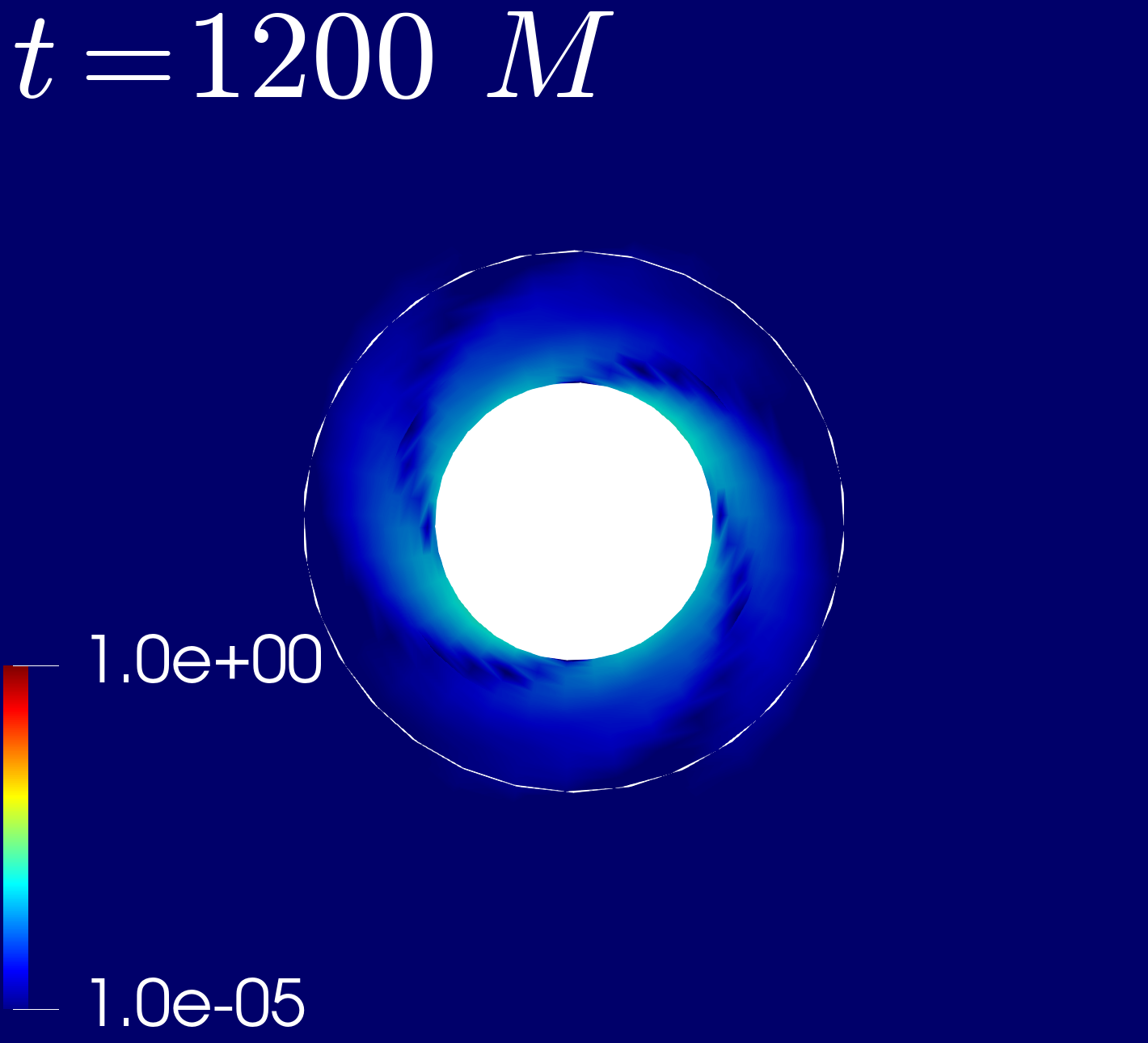} } \hspace{-0.15cm}
\subfloat{\includegraphics[width=0.24\textwidth]{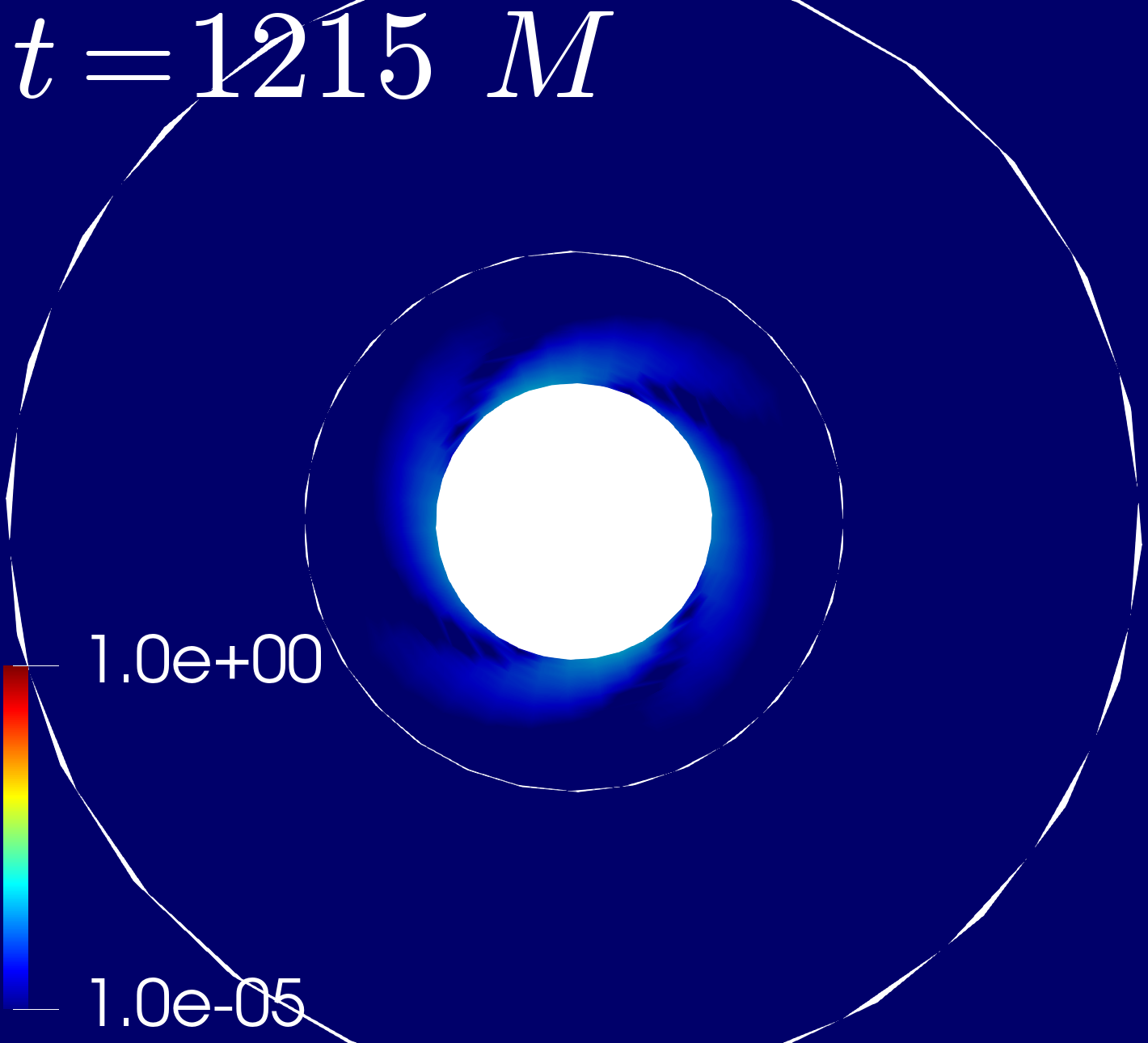} }\\
\caption{Similar to Fig.~\ref{fig:Speciality}, but for the \textbf{Type D1} Kerrness measure (cf.~\cite{Bhagwat:2017tkm}). Larger values correspond to more non-linearity, with a span of five orders of magnitude (the lower bound is set by the numerical noise floor as detailed in Sec.~\ref{sec:numerical_details}). We see the development of non-linearities close to merger, and though some are encompassed by the common horizon at its formation, the rest continue to enter the common horizon (and hence the event horizon) with time. The final state is a Kerr black hole. 
}
\label{fig:TypeD1}
\end{figure*}

\begin{figure*}
\subfloat{\includegraphics[width=0.24\textwidth]{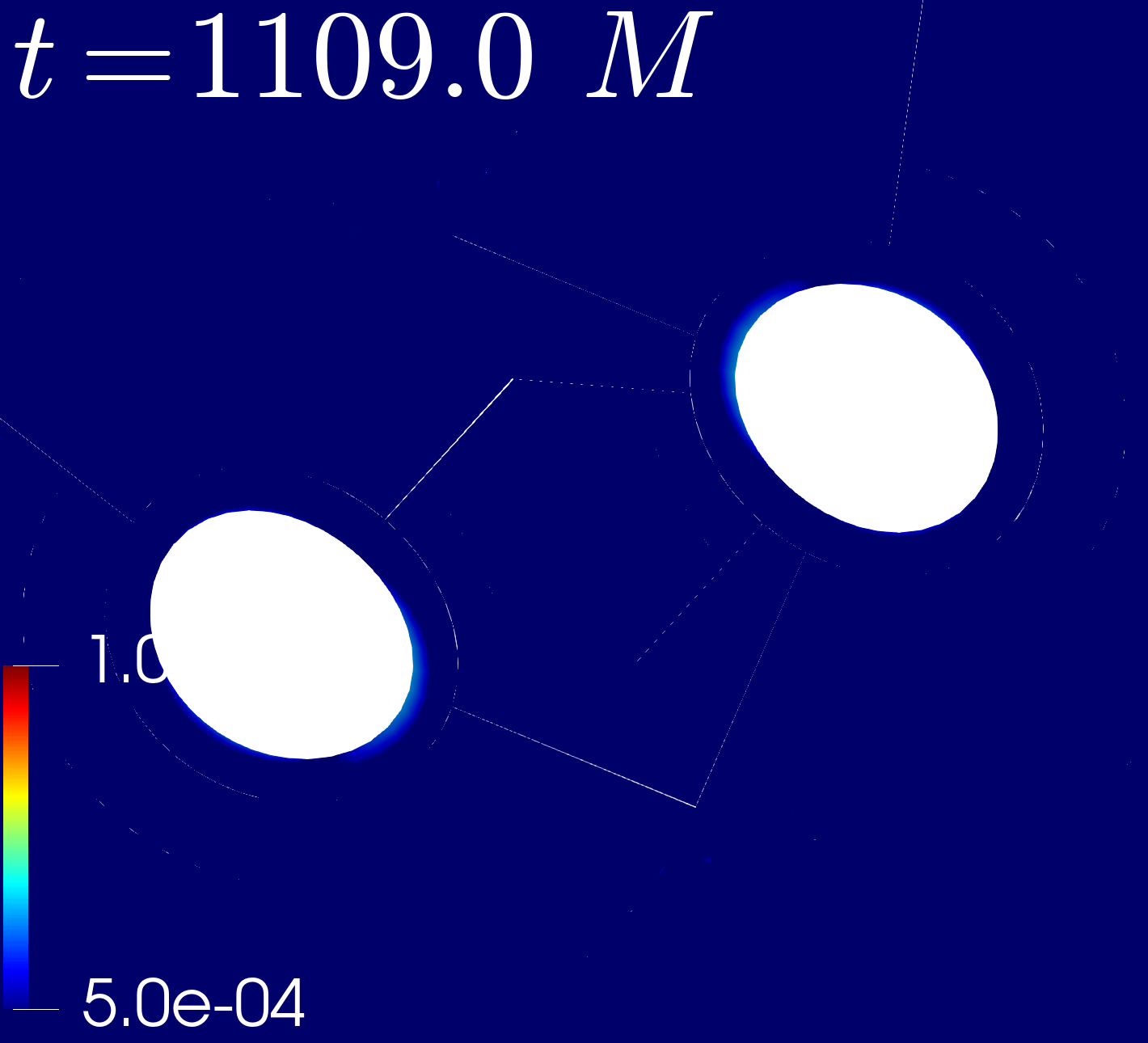} } \hspace{-0.15cm}
\subfloat{\includegraphics[width=0.24\textwidth]{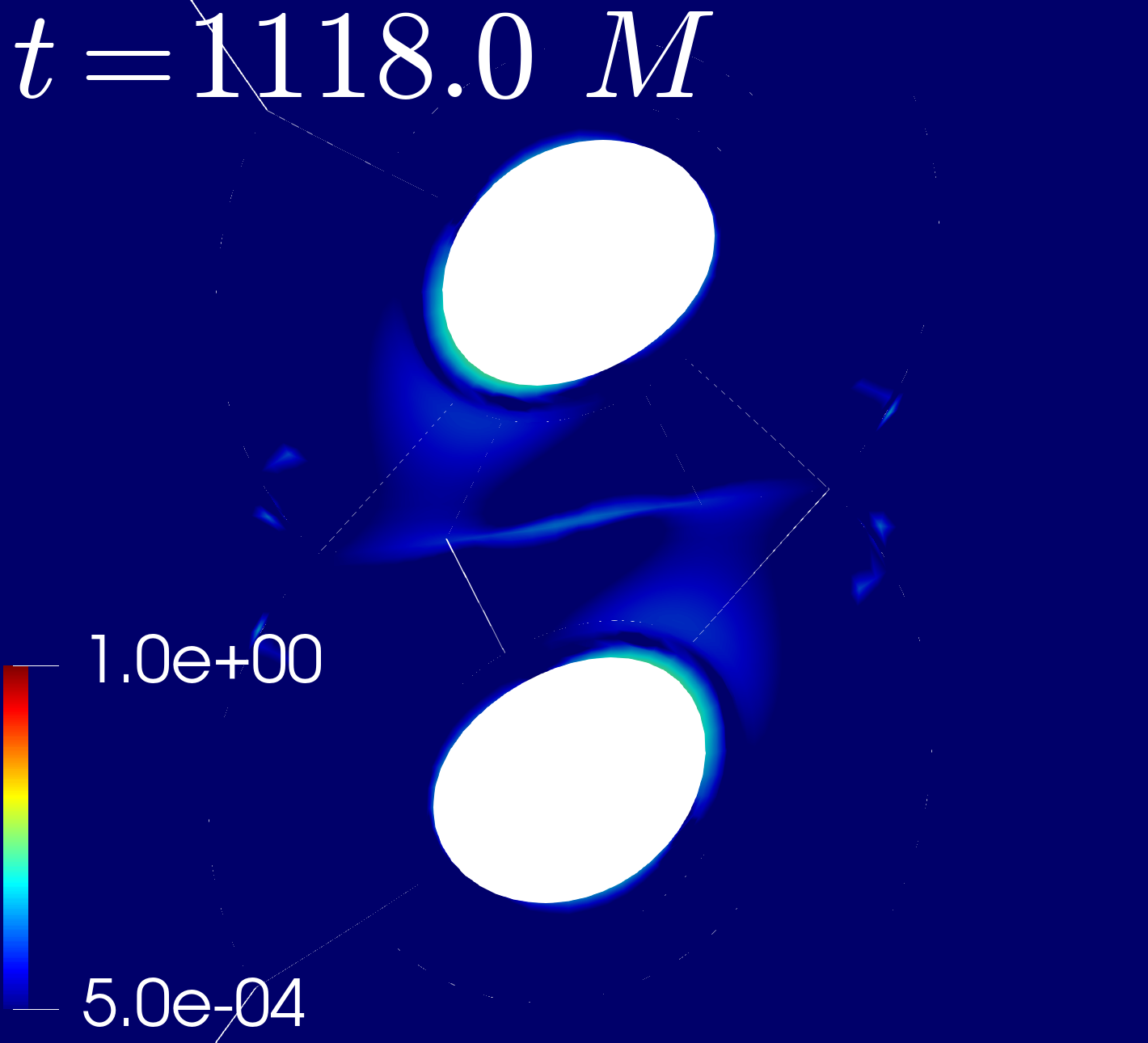} }  \hspace{-0.15cm}
\subfloat{\includegraphics[width=0.24\textwidth]{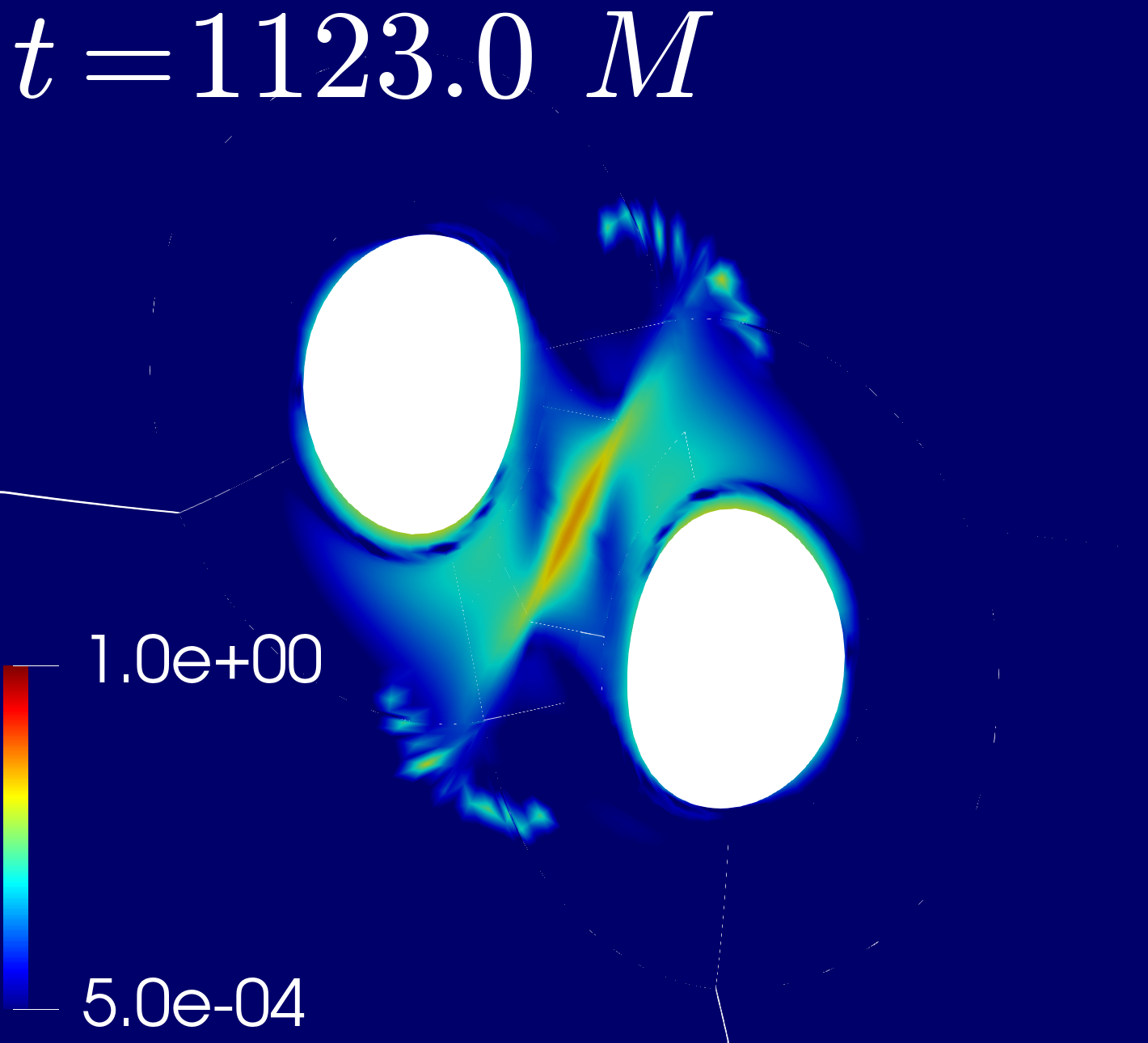} } \hspace{-0.15cm}
\subfloat{\includegraphics[width=0.24\textwidth]{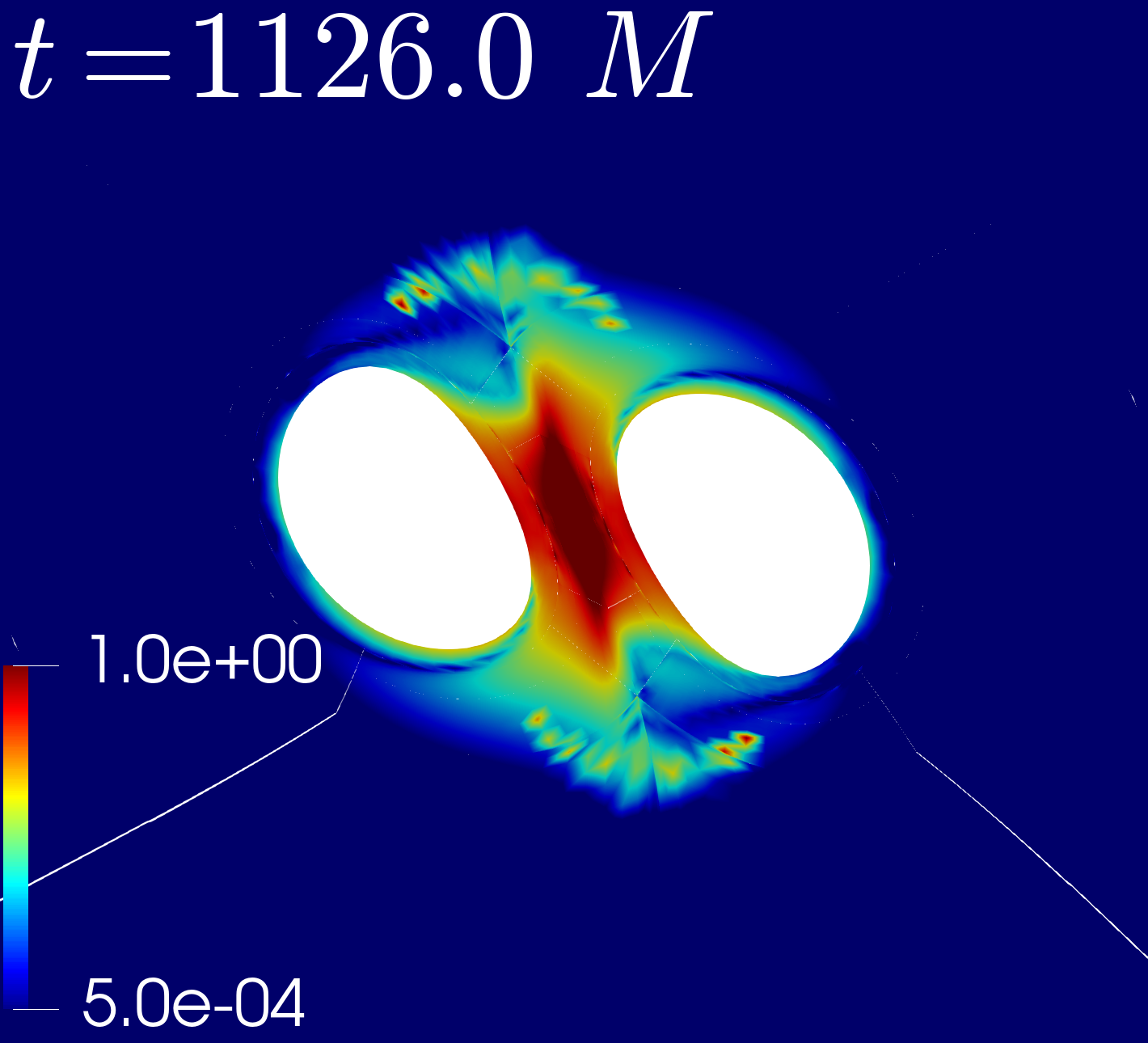} } \\ [-0.35cm]
\subfloat{\includegraphics[width=0.24\textwidth]{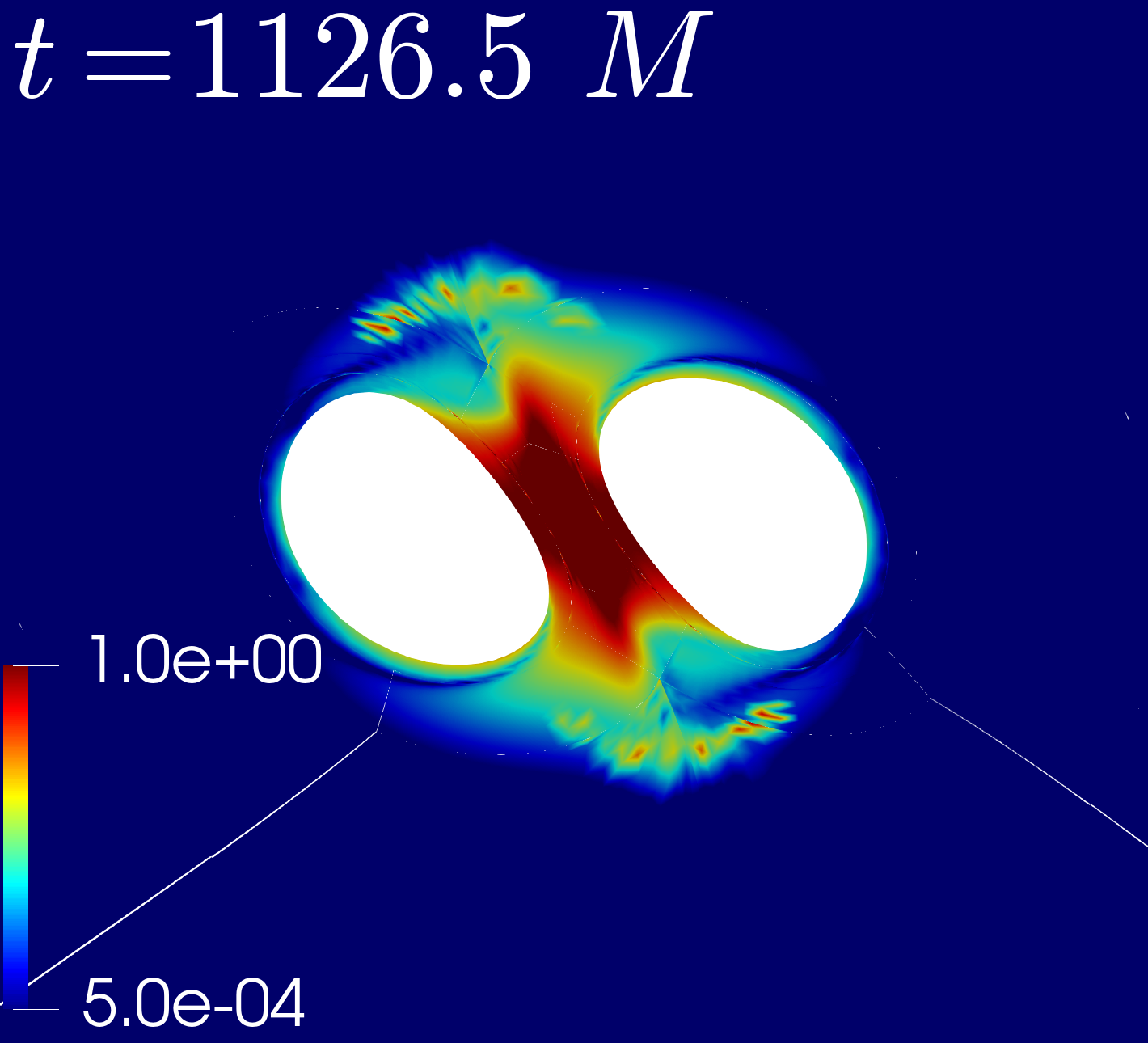} } \hspace{-0.15cm}
\subfloat{\includegraphics[width=0.24\textwidth]{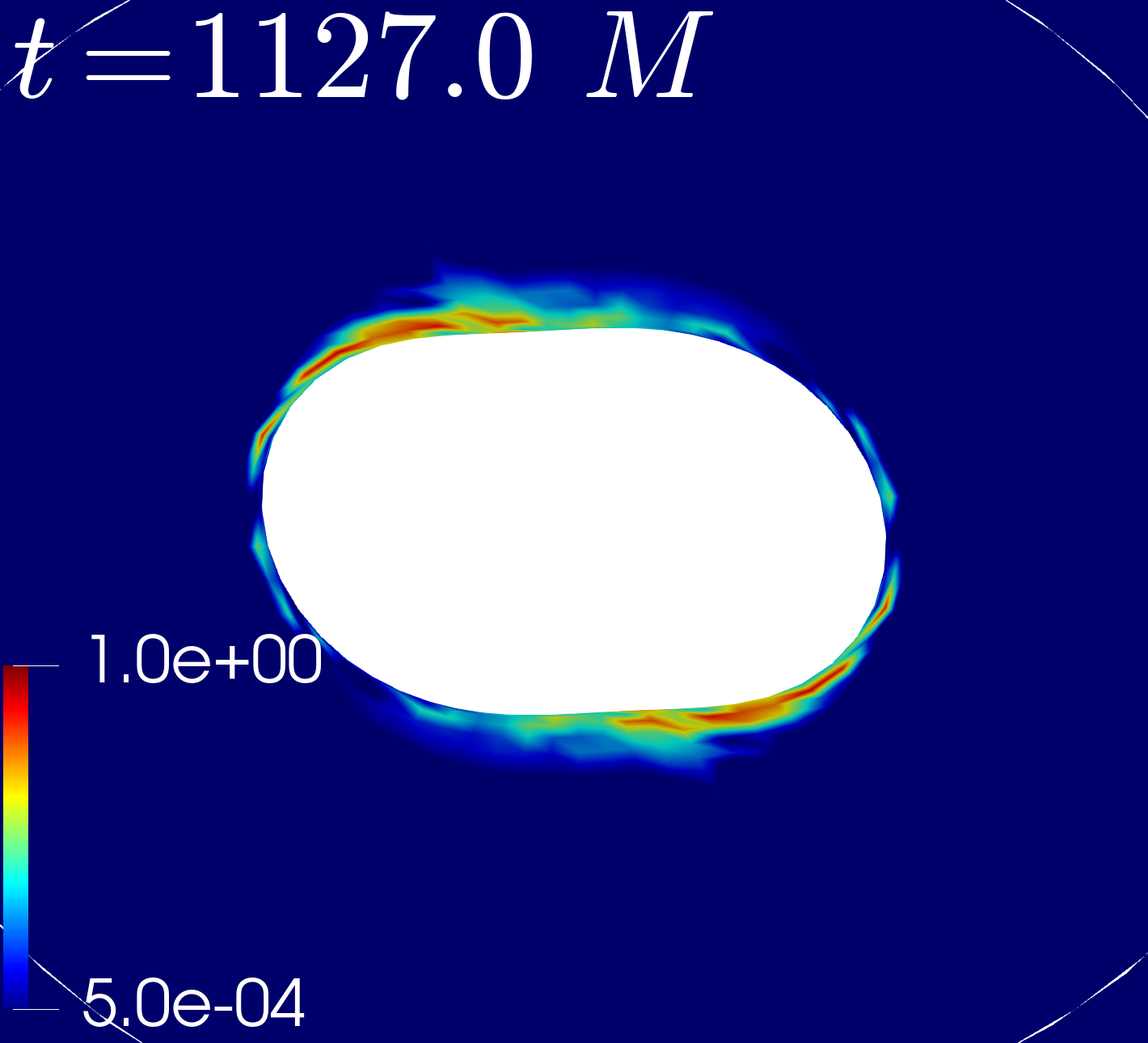} }  \hspace{-0.15cm}
\subfloat{\includegraphics[width=0.24\textwidth]{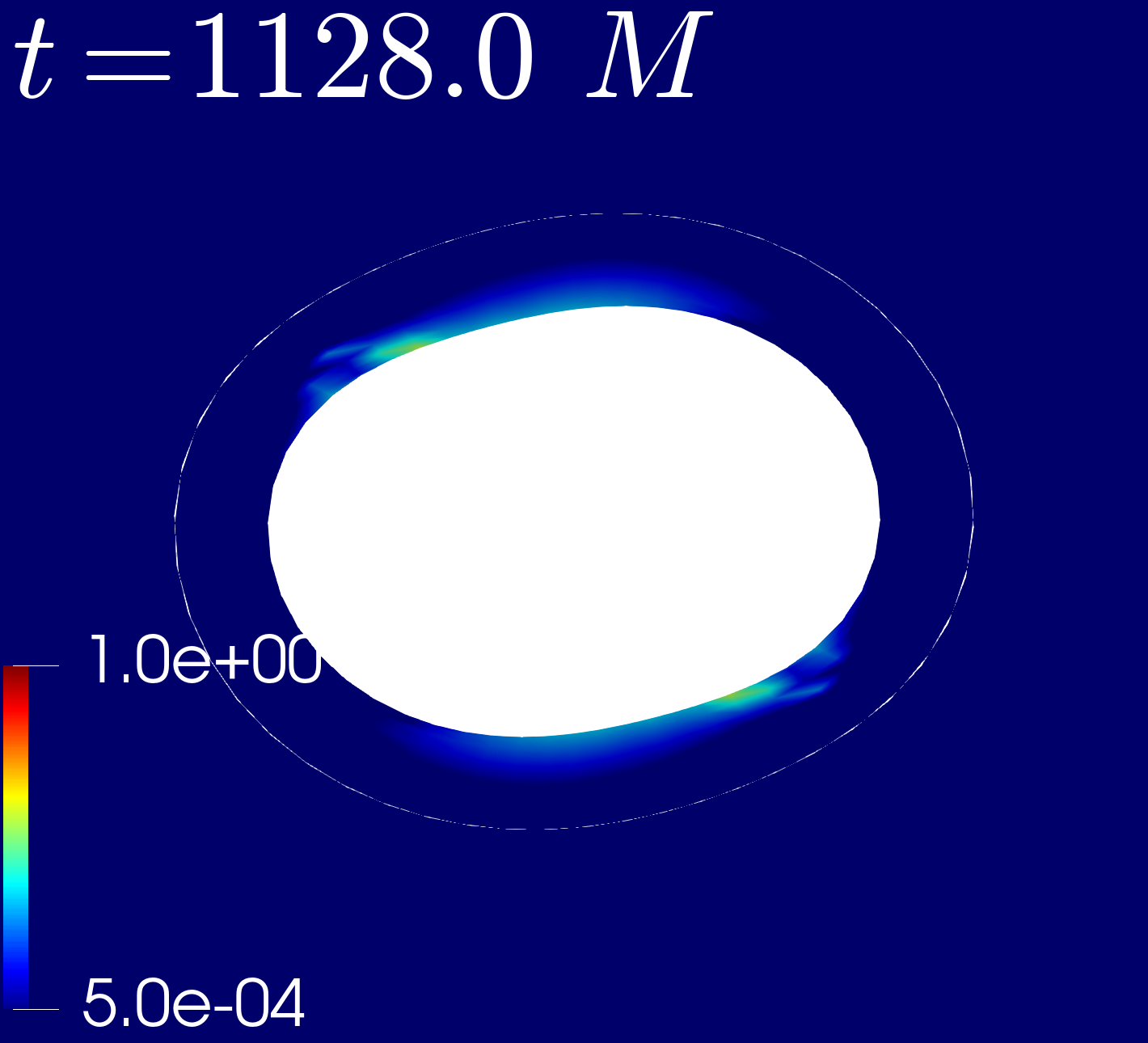} } \hspace{-0.15cm}
\subfloat{\includegraphics[width=0.24\textwidth]{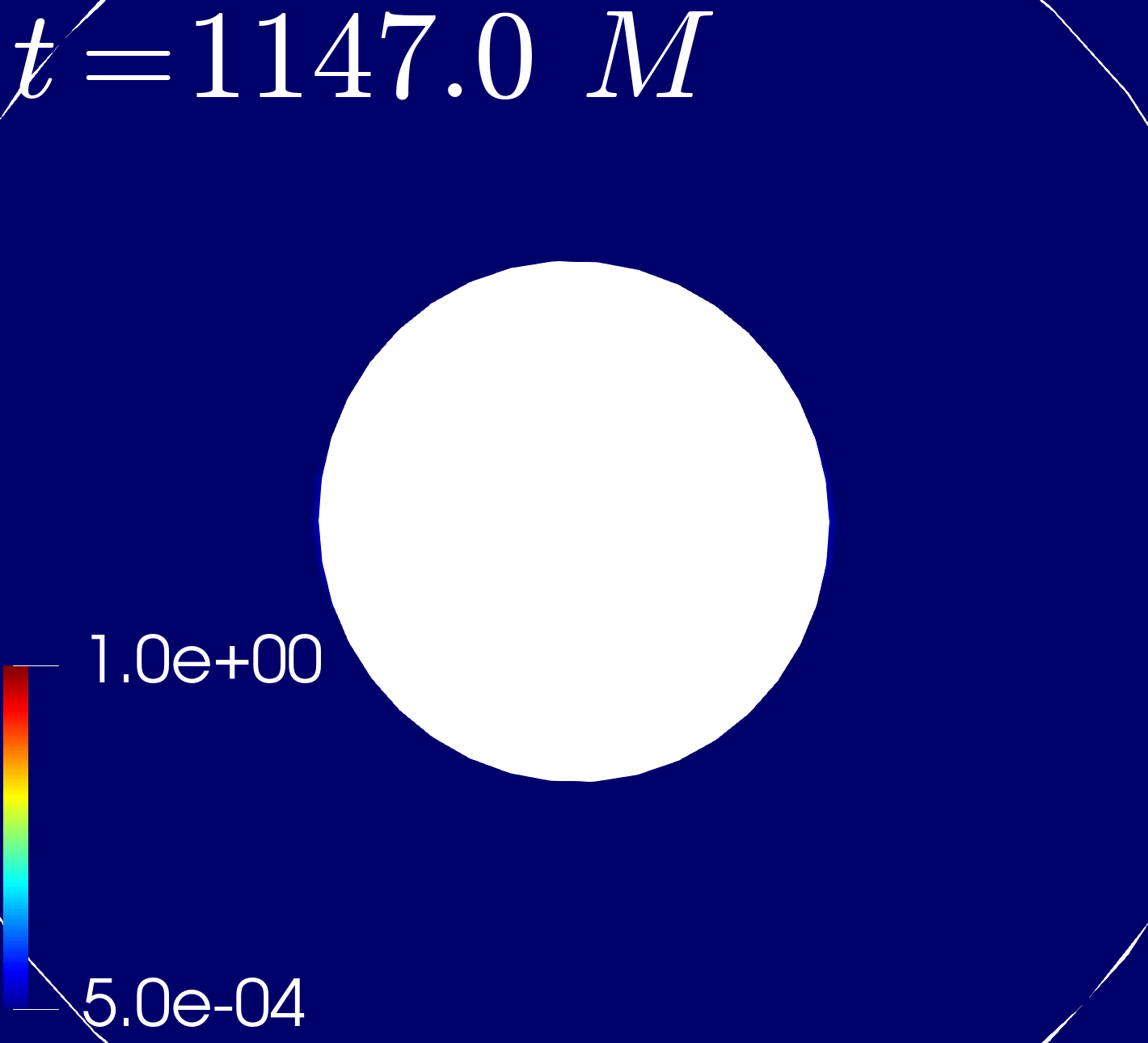} }
\caption{Similar to Fig.~\ref{fig:TypeD1}, but for the \textbf{Type D2} Kerrness measure (cf.~\cite{Bhagwat:2017tkm}). Note that compared to the \textbf{Type D1} measure presented in Fig.~\ref{fig:TypeD1}, this measure contains a third derivative of the spacetime metric (cf.~\cite{Bhagwat:2017tkm}), and thus contains more numerical noise, leading to some of the noisy artifacts (for example the horizontal line between the two black holes at $t = 1118\,M$ that corresponds to a domain boundary). We see that, as time progresses, the non-linearities are strongest between the two holes, and are mostly included inside the common horizon at the time of formation ($t = 1127.0\,M$). As time progresses, the remaining non-linearities continue to go down the common horizon (compare the $t = 1127.0\,M$ and $t = 1128.0\,M$ panels).}
\label{fig:TypeD2}
\end{figure*}

Next, in Figs.~\ref{fig:TypeD1} and~\ref{fig:TypeD2}, we show the \textbf{Type D1} and \textbf{Type D2} measures in the nomenclature of~\cite{Bhagwat:2017tkm}. These measures check whether the spacetime is Petrov type D (cf.~\cite{stephani2009exact}). As explained in Sec.~\ref{sec:methods}, we use these measures to quantify non-linearities in the spacetime. We see that for the Type D1 measure, non-linearities increase as the black holes merge, and though some of the initial non-linearities as quantified by Type D1 are immediately encompassed by the common horizon upon formation, the other ones go into the common horizon (and hence the event horizon) with time. A similar picture holds for Type D2. We see a quiescent Kerr black hole at late times in each case. 

Movies of these Kerrness measures evolving in time can be downloaded at

\href{https://github.com/mariaokounkova/BBHNonlinearity}{\texttt{github.com/mariaokounkova/BBHNonlinearity}}

\makeatletter\onecolumngrid@push\makeatother

\clearpage
\makeatletter\onecolumngrid@pop\makeatother

\makeatletter\onecolumngrid@push\makeatother

\clearpage
\makeatletter\onecolumngrid@pop\makeatother

\makeatletter\onecolumngrid@push\makeatother

\clearpage
\makeatletter\onecolumngrid@pop\makeatother

\section{Discussion and conclusion}
\label{sec:discussion}

In this work, we aimed to explore the recent results of Giesler et al.~\cite{Giesler:2019uxc}, which showed that the post-peak gravitational radiation in a numerical binary black hole merger can be fully described by linear perturbation theory. Given that binary black hole mergers are violent, non-linear events, the question arises of where the non-linearities in the spacetime `go' if they do not make it out to future null infinity, where gravitational detectors live. 

We investigated the behavior of non-linearities in the strong-field region of a numerical relativity binary black hole merger. We used a key result from~\cite{Bhagwat:2017tkm} that a set of \textit{Kerrness measures} can be used to quantify non-linearities in black hole spacetimes, as explained in Sec.~\ref{sec:methods}. Using these measures, we found that in the strong-field region, strong non-linearities develop close to merger, most of which are immediately encompassed by the common horizon (cf. Sec.~\ref{sec:horizons}), and the rest of which make their way into the common horizon with time. Since the common horizon lies inside of the event horizon, these non-linearities will not make it out to the gravitational wave detector. 

We also showed that the post-peak gravitational radiation associated with this simulation, as seen by a gravitational wave observer, is fully describable by linear theory (cf. Sec.~\ref{sec:overtones}), corroborating the results of~\cite{Giesler:2019uxc}.

This investigation, a numerical experiment, is one piece of one aspect of the puzzle brought on by the results of~\cite{Giesler:2019uxc}. There are many outstanding theoretical questions, including considering why the post-merger gravitational radiation is so well-described using the properties of the final black hole mass and spin, even when the post-merger black hole has a changing mass and angular momentum (cf.~\cite{Prasad:2020xgr} for one recent work on this topic). There are many more theoretical investigations to be done in this line of work, and perhaps this numerical experiment can be used to inform some of them.\footnote{Additionally, the equal-mass non-spinning simulation that we have considered in this paper (cf. Sec.~\ref{sec:simulation}) is a simple proof-of-concept case. Additional configurations, including high spins, for example, may be more interesting probes of non-linearity, and thus future work includes investigating the behavior of non-linearities in additional regions of binary black hole parameter space. }


\section*{Acknowledgements}

We thank Will Farr for helpful discussions. The Flatiron Institute is supported by the Simons Foundation. Computations were performed using the Spectral Einstein Code~\cite{SpECwebsite}. All computations were performed on the Wheeler cluster at Caltech, which is supported by the Sherman Fairchild Foundation and by Caltech.

\bibliography{biblio}

\begin{thebibliography}{31}%
\makeatletter
\providecommand \@ifxundefined [1]{%
 \@ifx{#1\undefined}
}%
\providecommand \@ifnum [1]{%
 \ifnum #1\expandafter \@firstoftwo
 \else \expandafter \@secondoftwo
 \fi
}%
\providecommand \@ifx [1]{%
 \ifx #1\expandafter \@firstoftwo
 \else \expandafter \@secondoftwo
 \fi
}%
\providecommand \natexlab [1]{#1}%
\providecommand \enquote  [1]{``#1''}%
\providecommand \bibnamefont  [1]{#1}%
\providecommand \bibfnamefont [1]{#1}%
\providecommand \citenamefont [1]{#1}%
\providecommand \href@noop [0]{\@secondoftwo}%
\providecommand \href [0]{\begingroup \@sanitize@url \@href}%
\providecommand \@href[1]{\@@startlink{#1}\@@href}%
\providecommand \@@href[1]{\endgroup#1\@@endlink}%
\providecommand \@sanitize@url [0]{\catcode `\\12\catcode `\$12\catcode
  `\&12\catcode `\#12\catcode `\^12\catcode `\_12\catcode `\%12\relax}%
\providecommand \@@startlink[1]{}%
\providecommand \@@endlink[0]{}%
\providecommand \url  [0]{\begingroup\@sanitize@url \@url }%
\providecommand \@url [1]{\endgroup\@href {#1}{\urlprefix }}%
\providecommand \urlprefix  [0]{URL }%
\providecommand \Eprint [0]{\href }%
\providecommand \doibase [0]{http://dx.doi.org/}%
\providecommand \selectlanguage [0]{\@gobble}%
\providecommand \bibinfo  [0]{\@secondoftwo}%
\providecommand \bibfield  [0]{\@secondoftwo}%
\providecommand \translation [1]{[#1]}%
\providecommand \BibitemOpen [0]{}%
\providecommand \bibitemStop [0]{}%
\providecommand \bibitemNoStop [0]{.\EOS\space}%
\providecommand \EOS [0]{\spacefactor3000\relax}%
\providecommand \BibitemShut  [1]{\csname bibitem#1\endcsname}%
\let\auto@bib@innerbib\@empty
\bibitem [{\citenamefont {Giesler}\ \emph {et~al.}(2019)\citenamefont
  {Giesler}, \citenamefont {Isi}, \citenamefont {Scheel},\ and\ \citenamefont
  {Teukolsky}}]{Giesler:2019uxc}%
  \BibitemOpen
  \bibfield  {author} {\bibinfo {author} {\bibfnamefont {M.}~\bibnamefont
  {Giesler}}, \bibinfo {author} {\bibfnamefont {M.}~\bibnamefont {Isi}},
  \bibinfo {author} {\bibfnamefont {M.}~\bibnamefont {Scheel}}, \ and\ \bibinfo
  {author} {\bibfnamefont {S.}~\bibnamefont {Teukolsky}},\ }\href@noop {} {\
  (\bibinfo {year} {2019})},\ \Eprint {http://arxiv.org/abs/1903.08284}
  {arXiv:1903.08284 [gr-qc]} \BibitemShut {NoStop}%
\bibitem [{\citenamefont {Isi}\ \emph {et~al.}(2019)\citenamefont {Isi},
  \citenamefont {Giesler}, \citenamefont {Farr}, \citenamefont {Scheel},\ and\
  \citenamefont {Teukolsky}}]{Isi:2019aib}%
  \BibitemOpen
  \bibfield  {author} {\bibinfo {author} {\bibfnamefont {M.}~\bibnamefont
  {Isi}}, \bibinfo {author} {\bibfnamefont {M.}~\bibnamefont {Giesler}},
  \bibinfo {author} {\bibfnamefont {W.~M.}\ \bibnamefont {Farr}}, \bibinfo
  {author} {\bibfnamefont {M.~A.}\ \bibnamefont {Scheel}}, \ and\ \bibinfo
  {author} {\bibfnamefont {S.~A.}\ \bibnamefont {Teukolsky}},\ }\href {\doibase
  10.1103/PhysRevLett.123.111102} {\bibfield  {journal} {\bibinfo  {journal}
  {Phys. Rev. Lett.}\ }\textbf {\bibinfo {volume} {123}},\ \bibinfo {pages}
  {111102} (\bibinfo {year} {2019})},\ \Eprint
  {http://arxiv.org/abs/1905.00869} {arXiv:1905.00869 [gr-qc]} \BibitemShut
  {NoStop}%
\bibitem [{\citenamefont {G\'omez-Lobo}(2016)}]{lobo16}%
  \BibitemOpen
  \bibfield  {author} {\bibinfo {author} {\bibfnamefont {A.~G.-P.}\
  \bibnamefont {G\'omez-Lobo}},\ }\href {\doibase
  10.1088/0264-9381/33/17/175005} {\bibfield  {journal} {\bibinfo  {journal}
  {Class. Quant. Grav.}\ }\textbf {\bibinfo {volume} {33}},\ \bibinfo {pages}
  {175005} (\bibinfo {year} {2016})},\ \Eprint
  {http://arxiv.org/abs/1602.08075} {arXiv:1602.08075 [gr-qc]} \BibitemShut
  {NoStop}%
\bibitem [{\citenamefont {Bhagwat}\ \emph {et~al.}(2018)\citenamefont
  {Bhagwat}, \citenamefont {Okounkova}, \citenamefont {Ballmer}, \citenamefont
  {Brown}, \citenamefont {Giesler}, \citenamefont {Scheel},\ and\ \citenamefont
  {Teukolsky}}]{Bhagwat:2017tkm}%
  \BibitemOpen
  \bibfield  {author} {\bibinfo {author} {\bibfnamefont {S.}~\bibnamefont
  {Bhagwat}}, \bibinfo {author} {\bibfnamefont {M.}~\bibnamefont {Okounkova}},
  \bibinfo {author} {\bibfnamefont {S.~W.}\ \bibnamefont {Ballmer}}, \bibinfo
  {author} {\bibfnamefont {D.~A.}\ \bibnamefont {Brown}}, \bibinfo {author}
  {\bibfnamefont {M.}~\bibnamefont {Giesler}}, \bibinfo {author} {\bibfnamefont
  {M.~A.}\ \bibnamefont {Scheel}}, \ and\ \bibinfo {author} {\bibfnamefont
  {S.~A.}\ \bibnamefont {Teukolsky}},\ }\href {\doibase
  10.1103/PhysRevD.97.104065} {\bibfield  {journal} {\bibinfo  {journal} {Phys.
  Rev.}\ }\textbf {\bibinfo {volume} {D97}},\ \bibinfo {pages} {104065}
  (\bibinfo {year} {2018})},\ \Eprint {http://arxiv.org/abs/1711.00926}
  {arXiv:1711.00926 [gr-qc]} \BibitemShut {NoStop}%
\bibitem [{\citenamefont {Boyle}\ and\ \citenamefont
  {Mroue}(2009)}]{Boyle:2009vi}%
  \BibitemOpen
  \bibfield  {author} {\bibinfo {author} {\bibfnamefont {M.}~\bibnamefont
  {Boyle}}\ and\ \bibinfo {author} {\bibfnamefont {A.~H.}\ \bibnamefont
  {Mroue}},\ }\href {\doibase 10.1103/PhysRevD.80.124045} {\bibfield  {journal}
  {\bibinfo  {journal} {Phys. Rev.}\ }\textbf {\bibinfo {volume} {D80}},\
  \bibinfo {pages} {124045} (\bibinfo {year} {2009})},\ \Eprint
  {http://arxiv.org/abs/0905.3177} {arXiv:0905.3177 [gr-qc]} \BibitemShut
  {NoStop}%
\bibitem [{\citenamefont {Baumgarte}\ and\ \citenamefont
  {Shapiro}(2010)}]{baumgarteShapiroBook}%
  \BibitemOpen
  \bibfield  {author} {\bibinfo {author} {\bibfnamefont {T.~W.}\ \bibnamefont
  {Baumgarte}}\ and\ \bibinfo {author} {\bibfnamefont {S.~L.}\ \bibnamefont
  {Shapiro}},\ }\href@noop {} {\emph {\bibinfo {title} {Numerical Relativity:
  Solving Einstein's Equations on the Computer}}}\ (\bibinfo  {publisher}
  {Cambridge University Press},\ \bibinfo {address} {Cambridge, England},\
  \bibinfo {year} {2010})\BibitemShut {NoStop}%
\bibitem [{\citenamefont {Yang}\ \emph {et~al.}(2015)\citenamefont {Yang},
  \citenamefont {Zimmerman},\ and\ \citenamefont {Lehner}}]{Yang:2014tla}%
  \BibitemOpen
  \bibfield  {author} {\bibinfo {author} {\bibfnamefont {H.}~\bibnamefont
  {Yang}}, \bibinfo {author} {\bibfnamefont {A.}~\bibnamefont {Zimmerman}}, \
  and\ \bibinfo {author} {\bibfnamefont {L.}~\bibnamefont {Lehner}},\ }\href
  {\doibase 10.1103/PhysRevLett.114.081101} {\bibfield  {journal} {\bibinfo
  {journal} {Phys. Rev. Lett.}\ }\textbf {\bibinfo {volume} {114}},\ \bibinfo
  {pages} {081101} (\bibinfo {year} {2015})},\ \Eprint
  {http://arxiv.org/abs/1402.4859} {arXiv:1402.4859 [gr-qc]} \BibitemShut
  {NoStop}%
\bibitem [{\citenamefont {Lousto}\ and\ \citenamefont
  {Whiting}(2002)}]{Lousto:2002em}%
  \BibitemOpen
  \bibfield  {author} {\bibinfo {author} {\bibfnamefont {C.~O.}\ \bibnamefont
  {Lousto}}\ and\ \bibinfo {author} {\bibfnamefont {B.~F.}\ \bibnamefont
  {Whiting}},\ }\href {\doibase 10.1103/PhysRevD.66.024026} {\bibfield
  {journal} {\bibinfo  {journal} {Phys. Rev.}\ }\textbf {\bibinfo {volume}
  {D66}},\ \bibinfo {pages} {024026} (\bibinfo {year} {2002})},\ \Eprint
  {http://arxiv.org/abs/gr-qc/0203061} {arXiv:gr-qc/0203061 [gr-qc]}
  \BibitemShut {NoStop}%
\bibitem [{\citenamefont {Teukolsky}(2015)}]{Teukolsky:2014vca}%
  \BibitemOpen
  \bibfield  {author} {\bibinfo {author} {\bibfnamefont {S.~A.}\ \bibnamefont
  {Teukolsky}},\ }\href {\doibase 10.1088/0264-9381/32/12/124006} {\bibfield
  {journal} {\bibinfo  {journal} {Class. Quant. Grav.}\ }\textbf {\bibinfo
  {volume} {32}},\ \bibinfo {pages} {124006} (\bibinfo {year} {2015})},\
  \Eprint {http://arxiv.org/abs/1410.2130} {arXiv:1410.2130 [gr-qc]}
  \BibitemShut {NoStop}%
\bibitem [{\citenamefont {Cook}\ and\ \citenamefont
  {Pfeiffer}(2004)}]{Cook2004}%
  \BibitemOpen
  \bibfield  {author} {\bibinfo {author} {\bibfnamefont {G.~B.}\ \bibnamefont
  {Cook}}\ and\ \bibinfo {author} {\bibfnamefont {H.~P.}\ \bibnamefont
  {Pfeiffer}},\ }\href {\doibase 10.1103/PhysRevD.70.104016} {\bibfield
  {journal} {\bibinfo  {journal} {Phys. Rev. D}\ }\textbf {\bibinfo {volume}
  {70}},\ \bibinfo {pages} {104016} (\bibinfo {year} {2004})}\BibitemShut
  {NoStop}%
\bibitem [{\citenamefont {Pfeiffer}(2005)}]{Pfeiffer:2005zm}%
  \BibitemOpen
  \bibfield  {author} {\bibinfo {author} {\bibfnamefont {H.~P.}\ \bibnamefont
  {Pfeiffer}},\ }\emph {\bibinfo {title} {{Initial data for black hole
  evolutions}}},\ \href {http://wwwlib.umi.com/dissertations/fullcit?p3104429}
  {Ph.D. thesis},\ \bibinfo  {school} {Cornell U., Phys. Dept.} (\bibinfo
  {year} {2005}),\ \Eprint {http://arxiv.org/abs/gr-qc/0510016}
  {arXiv:gr-qc/0510016 [gr-qc]} \BibitemShut {NoStop}%
\bibitem [{\citenamefont {Lovelace}(2009)}]{Lovelace:2008hd}%
  \BibitemOpen
  \bibfield  {author} {\bibinfo {author} {\bibfnamefont {G.}~\bibnamefont
  {Lovelace}},\ }\bibfield  {booktitle} {\emph {\bibinfo {booktitle}
  {{Numerical relativity data analysis. Proceedings, 2nd Meeting, NRDA 2008,
  Syracuse, USA, August 11-14, 2008}}},\ }\href {\doibase
  10.1088/0264-9381/26/11/114002} {\bibfield  {journal} {\bibinfo  {journal}
  {Class. Quant. Grav.}\ }\textbf {\bibinfo {volume} {26}},\ \bibinfo {pages}
  {114002} (\bibinfo {year} {2009})},\ \Eprint {http://arxiv.org/abs/0812.3132}
  {arXiv:0812.3132 [gr-qc]} \BibitemShut {NoStop}%
\bibitem [{\citenamefont {Owen}(2009)}]{Owen:2009sb}%
  \BibitemOpen
  \bibfield  {author} {\bibinfo {author} {\bibfnamefont {R.}~\bibnamefont
  {Owen}},\ }\href {\doibase 10.1103/PhysRevD.80.084012} {\bibfield  {journal}
  {\bibinfo  {journal} {Phys. Rev.}\ }\textbf {\bibinfo {volume} {D80}},\
  \bibinfo {pages} {084012} (\bibinfo {year} {2009})},\ \Eprint
  {http://arxiv.org/abs/0907.0280} {arXiv:0907.0280 [gr-qc]} \BibitemShut
  {NoStop}%
\bibitem [{\citenamefont {Campanelli}\ \emph {et~al.}(2009)\citenamefont
  {Campanelli}, \citenamefont {Lousto},\ and\ \citenamefont
  {Zlochower}}]{Campanelli:2008dv}%
  \BibitemOpen
  \bibfield  {author} {\bibinfo {author} {\bibfnamefont {M.}~\bibnamefont
  {Campanelli}}, \bibinfo {author} {\bibfnamefont {C.~O.}\ \bibnamefont
  {Lousto}}, \ and\ \bibinfo {author} {\bibfnamefont {Y.}~\bibnamefont
  {Zlochower}},\ }\href {\doibase 10.1103/PhysRevD.79.084012} {\bibfield
  {journal} {\bibinfo  {journal} {Phys. Rev.}\ }\textbf {\bibinfo {volume}
  {D79}},\ \bibinfo {pages} {084012} (\bibinfo {year} {2009})},\ \Eprint
  {http://arxiv.org/abs/0811.3006} {arXiv:0811.3006 [gr-qc]} \BibitemShut
  {NoStop}%
\bibitem [{\citenamefont {Wald}(1984)}]{Wald:106274}%
  \BibitemOpen
  \bibfield  {author} {\bibinfo {author} {\bibfnamefont {R.~M.}\ \bibnamefont
  {Wald}},\ }\href {https://cds.cern.ch/record/106274} {\emph {\bibinfo {title}
  {{General relativity}}}}\ (\bibinfo  {publisher} {Chicago Univ. Press},\
  \bibinfo {address} {Chicago, IL},\ \bibinfo {year} {1984})\BibitemShut
  {NoStop}%
\bibitem [{\citenamefont {Cohen}\ \emph {et~al.}(2009)\citenamefont {Cohen},
  \citenamefont {Pfeiffer},\ and\ \citenamefont {Scheel}}]{Cohen:2008wa}%
  \BibitemOpen
  \bibfield  {author} {\bibinfo {author} {\bibfnamefont {M.~I.}\ \bibnamefont
  {Cohen}}, \bibinfo {author} {\bibfnamefont {H.~P.}\ \bibnamefont {Pfeiffer}},
  \ and\ \bibinfo {author} {\bibfnamefont {M.~A.}\ \bibnamefont {Scheel}},\
  }\href {\doibase 10.1088/0264-9381/26/3/035005} {\bibfield  {journal}
  {\bibinfo  {journal} {Class. Quant. Grav.}\ }\textbf {\bibinfo {volume}
  {26}},\ \bibinfo {pages} {035005} (\bibinfo {year} {2009})},\ \Eprint
  {http://arxiv.org/abs/0809.2628} {arXiv:0809.2628 [gr-qc]} \BibitemShut
  {NoStop}%
\bibitem [{\citenamefont {Bohn}\ \emph
  {et~al.}(2016{\natexlab{a}})\citenamefont {Bohn}, \citenamefont {Kidder},\
  and\ \citenamefont {Teukolsky}}]{Bohn:2016soe}%
  \BibitemOpen
  \bibfield  {author} {\bibinfo {author} {\bibfnamefont {A.}~\bibnamefont
  {Bohn}}, \bibinfo {author} {\bibfnamefont {L.~E.}\ \bibnamefont {Kidder}}, \
  and\ \bibinfo {author} {\bibfnamefont {S.~A.}\ \bibnamefont {Teukolsky}},\
  }\href {\doibase 10.1103/PhysRevD.94.064009} {\bibfield  {journal} {\bibinfo
  {journal} {Phys. Rev.}\ }\textbf {\bibinfo {volume} {D94}},\ \bibinfo {pages}
  {064009} (\bibinfo {year} {2016}{\natexlab{a}})},\ \Eprint
  {http://arxiv.org/abs/1606.00436} {arXiv:1606.00436 [gr-qc]} \BibitemShut
  {NoStop}%
\bibitem [{\citenamefont {Bohn}\ \emph
  {et~al.}(2016{\natexlab{b}})\citenamefont {Bohn}, \citenamefont {Kidder},\
  and\ \citenamefont {Teukolsky}}]{Bohn:2016afc}%
  \BibitemOpen
  \bibfield  {author} {\bibinfo {author} {\bibfnamefont {A.}~\bibnamefont
  {Bohn}}, \bibinfo {author} {\bibfnamefont {L.~E.}\ \bibnamefont {Kidder}}, \
  and\ \bibinfo {author} {\bibfnamefont {S.~A.}\ \bibnamefont {Teukolsky}},\
  }\href {\doibase 10.1103/PhysRevD.94.064008} {\bibfield  {journal} {\bibinfo
  {journal} {Phys. Rev.}\ }\textbf {\bibinfo {volume} {D94}},\ \bibinfo {pages}
  {064008} (\bibinfo {year} {2016}{\natexlab{b}})},\ \Eprint
  {http://arxiv.org/abs/1606.00437} {arXiv:1606.00437 [gr-qc]} \BibitemShut
  {NoStop}%
\bibitem [{\citenamefont {Thornburg}(1996)}]{Thornburg:1995cp}%
  \BibitemOpen
  \bibfield  {author} {\bibinfo {author} {\bibfnamefont {J.}~\bibnamefont
  {Thornburg}},\ }\href {\doibase 10.1103/PhysRevD.54.4899} {\bibfield
  {journal} {\bibinfo  {journal} {Phys. Rev.}\ }\textbf {\bibinfo {volume}
  {D54}},\ \bibinfo {pages} {4899} (\bibinfo {year} {1996})},\ \Eprint
  {http://arxiv.org/abs/gr-qc/9508014} {arXiv:gr-qc/9508014 [gr-qc]}
  \BibitemShut {NoStop}%
\bibitem [{\citenamefont {Gundlach}(1998)}]{Gundlach:1997us}%
  \BibitemOpen
  \bibfield  {author} {\bibinfo {author} {\bibfnamefont {C.}~\bibnamefont
  {Gundlach}},\ }\href {\doibase 10.1103/PhysRevD.57.863} {\bibfield  {journal}
  {\bibinfo  {journal} {Phys. Rev.}\ }\textbf {\bibinfo {volume} {D57}},\
  \bibinfo {pages} {863} (\bibinfo {year} {1998})},\ \Eprint
  {http://arxiv.org/abs/gr-qc/9707050} {arXiv:gr-qc/9707050 [gr-qc]}
  \BibitemShut {NoStop}%
\bibitem [{\citenamefont {Lovelace}\ \emph {et~al.}(2015)\citenamefont
  {Lovelace} \emph {et~al.}}]{Lovelace:2014twa}%
  \BibitemOpen
  \bibfield  {author} {\bibinfo {author} {\bibfnamefont {G.}~\bibnamefont
  {Lovelace}} \emph {et~al.},\ }\href {\doibase 10.1088/0264-9381/32/6/065007}
  {\bibfield  {journal} {\bibinfo  {journal} {Class. Quant. Grav.}\ }\textbf
  {\bibinfo {volume} {32}},\ \bibinfo {pages} {065007} (\bibinfo {year}
  {2015})},\ \Eprint {http://arxiv.org/abs/1411.7297} {arXiv:1411.7297 [gr-qc]}
  \BibitemShut {NoStop}%
\bibitem [{SpE()}]{SpECwebsite}%
  \BibitemOpen
  \href@noop {} {\enquote {\bibinfo {title} {The {S}pectral {E}instein {C}ode
  ({SpEC})},}\ }\bibinfo {howpublished}
  {\url{http://www.black-holes.org/SpEC.html}}\BibitemShut {NoStop}%
\bibitem [{\citenamefont {Hemberger}\ \emph {et~al.}(2013)\citenamefont
  {Hemberger}, \citenamefont {Scheel}, \citenamefont {Kidder}, \citenamefont
  {Szil{\'a}gyi}, \citenamefont {Lovelace}, \citenamefont {Taylor},\ and\
  \citenamefont {Teukolsky}}]{Hemberger:2012jz}%
  \BibitemOpen
  \bibfield  {author} {\bibinfo {author} {\bibfnamefont {D.~A.}\ \bibnamefont
  {Hemberger}}, \bibinfo {author} {\bibfnamefont {M.~A.}\ \bibnamefont
  {Scheel}}, \bibinfo {author} {\bibfnamefont {L.~E.}\ \bibnamefont {Kidder}},
  \bibinfo {author} {\bibfnamefont {B.}~\bibnamefont {Szil{\'a}gyi}}, \bibinfo
  {author} {\bibfnamefont {G.}~\bibnamefont {Lovelace}}, \bibinfo {author}
  {\bibfnamefont {N.~W.}\ \bibnamefont {Taylor}}, \ and\ \bibinfo {author}
  {\bibfnamefont {S.~A.}\ \bibnamefont {Teukolsky}},\ }\href {\doibase
  10.1088/0264-9381/30/11/115001} {\bibfield  {journal} {\bibinfo  {journal}
  {Class. Quant. Grav.}\ }\textbf {\bibinfo {volume} {30}},\ \bibinfo {pages}
  {115001} (\bibinfo {year} {2013})},\ \Eprint {http://arxiv.org/abs/1211.6079}
  {arXiv:1211.6079 [gr-qc]} \BibitemShut {NoStop}%
\bibitem [{SXS()}]{SXSCatalog}%
  \BibitemOpen
  \href@noop {} {\enquote {\bibinfo {title} {Sxs gravitational waveform
  database},}\ }\bibinfo {howpublished}
  {\url{http://www.black-holes.org/waveforms}}\BibitemShut {NoStop}%
\bibitem [{\citenamefont {Lindblom}\ \emph {et~al.}(2006)\citenamefont
  {Lindblom}, \citenamefont {Scheel}, \citenamefont {Kidder}, \citenamefont
  {Owen},\ and\ \citenamefont {Rinne}}]{Lindblom2006}%
  \BibitemOpen
  \bibfield  {author} {\bibinfo {author} {\bibfnamefont {L.}~\bibnamefont
  {Lindblom}}, \bibinfo {author} {\bibfnamefont {M.~A.}\ \bibnamefont
  {Scheel}}, \bibinfo {author} {\bibfnamefont {L.~E.}\ \bibnamefont {Kidder}},
  \bibinfo {author} {\bibfnamefont {R.}~\bibnamefont {Owen}}, \ and\ \bibinfo
  {author} {\bibfnamefont {O.}~\bibnamefont {Rinne}},\ }\href {\doibase
  10.1088/0264-9381/23/16/S09} {\bibfield  {journal} {\bibinfo  {journal}
  {Class. Quant. Grav.}\ }\textbf {\bibinfo {volume} {23}},\ \bibinfo {pages}
  {S447} (\bibinfo {year} {2006})},\ \Eprint
  {http://arxiv.org/abs/gr-qc/0512093} {arXiv:gr-qc/0512093 [gr-qc]}
  \BibitemShut {NoStop}%
\bibitem [{\citenamefont {Scheel}\ \emph {et~al.}(2009)\citenamefont {Scheel},
  \citenamefont {Boyle}, \citenamefont {Chu}, \citenamefont {Kidder},
  \citenamefont {Matthews},\ and\ \citenamefont {Pfeiffer}}]{Scheel:2008rj}%
  \BibitemOpen
  \bibfield  {author} {\bibinfo {author} {\bibfnamefont {M.~A.}\ \bibnamefont
  {Scheel}}, \bibinfo {author} {\bibfnamefont {M.}~\bibnamefont {Boyle}},
  \bibinfo {author} {\bibfnamefont {T.}~\bibnamefont {Chu}}, \bibinfo {author}
  {\bibfnamefont {L.~E.}\ \bibnamefont {Kidder}}, \bibinfo {author}
  {\bibfnamefont {K.~D.}\ \bibnamefont {Matthews}}, \ and\ \bibinfo {author}
  {\bibfnamefont {H.~P.}\ \bibnamefont {Pfeiffer}},\ }\href {\doibase
  10.1103/PhysRevD.79.024003} {\bibfield  {journal} {\bibinfo  {journal} {Phys.
  Rev.}\ }\textbf {\bibinfo {volume} {D79}},\ \bibinfo {pages} {024003}
  (\bibinfo {year} {2009})},\ \Eprint {http://arxiv.org/abs/0810.1767}
  {arXiv:0810.1767 [gr-qc]} \BibitemShut {NoStop}%
\bibitem [{\citenamefont {Szilagyi}\ \emph {et~al.}(2009)\citenamefont
  {Szilagyi}, \citenamefont {Lindblom},\ and\ \citenamefont
  {Scheel}}]{Szilagyi:2009qz}%
  \BibitemOpen
  \bibfield  {author} {\bibinfo {author} {\bibfnamefont {B.}~\bibnamefont
  {Szilagyi}}, \bibinfo {author} {\bibfnamefont {L.}~\bibnamefont {Lindblom}},
  \ and\ \bibinfo {author} {\bibfnamefont {M.~A.}\ \bibnamefont {Scheel}},\
  }\href {\doibase 10.1103/PhysRevD.80.124010} {\bibfield  {journal} {\bibinfo
  {journal} {Phys. Rev.}\ }\textbf {\bibinfo {volume} {D80}},\ \bibinfo {pages}
  {124010} (\bibinfo {year} {2009})},\ \Eprint {http://arxiv.org/abs/0909.3557}
  {arXiv:0909.3557 [gr-qc]} \BibitemShut {NoStop}%
\bibitem [{\citenamefont {Okounkova}\ \emph {et~al.}(2019)\citenamefont
  {Okounkova}, \citenamefont {Stein}, \citenamefont {Scheel},\ and\
  \citenamefont {Teukolsky}}]{MashaHeadOn}%
  \BibitemOpen
  \bibfield  {author} {\bibinfo {author} {\bibfnamefont {M.}~\bibnamefont
  {Okounkova}}, \bibinfo {author} {\bibfnamefont {L.~C.}\ \bibnamefont
  {Stein}}, \bibinfo {author} {\bibfnamefont {M.~A.}\ \bibnamefont {Scheel}}, \
  and\ \bibinfo {author} {\bibfnamefont {S.~A.}\ \bibnamefont {Teukolsky}},\
  }\href@noop {} {\  (\bibinfo {year} {2019})},\ \Eprint
  {http://arxiv.org/abs/1906.08789} {arXiv:1906.08789 [gr-qc]} \BibitemShut
  {NoStop}%
\bibitem [{\citenamefont {Stein}(2019)}]{QNMCode}%
  \BibitemOpen
  \bibfield  {author} {\bibinfo {author} {\bibfnamefont {L.~C.}\ \bibnamefont
  {Stein}},\ }\href {\doibase 10.21105/joss.01683} {\bibfield  {journal}
  {\bibinfo  {journal} {J. Open Source Softw.}\ }\textbf {\bibinfo {volume}
  {4}},\ \bibinfo {pages} {1683} (\bibinfo {year} {2019})},\ \Eprint
  {http://arxiv.org/abs/1908.10377} {arXiv:1908.10377 [gr-qc]} \BibitemShut
  {NoStop}%
\bibitem [{\citenamefont {Stephani}\ \emph {et~al.}(2009)\citenamefont
  {Stephani}, \citenamefont {Kramer}, \citenamefont {MacCallum}, \citenamefont
  {Hoenselaers},\ and\ \citenamefont {Herlt}}]{stephani2009exact}%
  \BibitemOpen
  \bibfield  {author} {\bibinfo {author} {\bibfnamefont {H.}~\bibnamefont
  {Stephani}}, \bibinfo {author} {\bibfnamefont {D.}~\bibnamefont {Kramer}},
  \bibinfo {author} {\bibfnamefont {M.}~\bibnamefont {MacCallum}}, \bibinfo
  {author} {\bibfnamefont {C.}~\bibnamefont {Hoenselaers}}, \ and\ \bibinfo
  {author} {\bibfnamefont {E.}~\bibnamefont {Herlt}},\ }\href
  {https://books.google.com/books?id=SiWXP8FjTFEC} {\emph {\bibinfo {title}
  {Exact Solutions of Einstein's Field Equations}}},\ Cambridge Monographs on
  Mathematical Physics\ (\bibinfo  {publisher} {Cambridge University Press},\
  \bibinfo {year} {2009})\BibitemShut {NoStop}%
\bibitem [{\citenamefont {Prasad}\ \emph {et~al.}(2020)\citenamefont {Prasad},
  \citenamefont {Gupta}, \citenamefont {Bose}, \citenamefont {Krishnan},\ and\
  \citenamefont {Schnetter}}]{Prasad:2020xgr}%
  \BibitemOpen
  \bibfield  {author} {\bibinfo {author} {\bibfnamefont {V.}~\bibnamefont
  {Prasad}}, \bibinfo {author} {\bibfnamefont {A.}~\bibnamefont {Gupta}},
  \bibinfo {author} {\bibfnamefont {S.}~\bibnamefont {Bose}}, \bibinfo {author}
  {\bibfnamefont {B.}~\bibnamefont {Krishnan}}, \ and\ \bibinfo {author}
  {\bibfnamefont {E.}~\bibnamefont {Schnetter}},\ }\href@noop {} {\  (\bibinfo
  {year} {2020})},\ \Eprint {http://arxiv.org/abs/2003.06215} {arXiv:2003.06215
  [gr-qc]} \BibitemShut {NoStop}%
\end{thebibliography}%
\end{document}